\documentclass[manuscript,screen]{acmart}
\usepackage{CJKutf8}
\usepackage{ulem}
\usepackage{multirow}
\usepackage{makecell}
\usepackage{bbding}
\usepackage{hyperref} 
\usepackage{longtable}
\usepackage[table]{xcolor}

\AtBeginDocument{%
  }

\setcopyright{acmlicensed}
\copyrightyear{2025}
\acmYear{2025}
\acmDOI{XXXXXXX.XXXXXXX}
\acmConference[Conference acronym 'XX]{Make sure to enter the correct
  conference title from your rights confirmation email}{June 03--05,
  2018}{Woodstock, NY}
\acmISBN{978-1-4503-XXXX-X/2018/06}

\begin{document}

\title{A Comprehensive Survey on Generative AI for Video-to-Music Generation}

\author{Shulei Ji}
\email{shuleiji@zju.edu.cn}
\orcid{0000-0002-9908-136X}
\affiliation{%
  \institution{Zhejiang University and Innovation Center of Yangtze River Delta, Zhejiang University}
  \city{Hangzhou}
  \state{Zhejiang}
  \country{China}
}

\author{Songruoyao Wu}
\affiliation{%
  \institution{Zhejiang University}
  \city{Hangzhou}
  \state{Zhejiang}
  \country{China}}
\email{wsry@zju.edu.cn}

\author{Zihao Wang}
\affiliation{%
  \institution{Zhejiang University}
  \city{Hangzhou}
  \state{Zhejiang}
  \country{China}
}
\affiliation{%
  \institution{Carnegie Mellon University}
  \country{USA}
}
\email{carlwang@zju.edu.cn}

\author{Shuyu Li}
\affiliation{%
  \institution{Zhejiang University}
  \city{Hangzhou}
  \state{Zhejiang}
  \country{China}
}
\email{lsyxary@zju.edu.cn}

\author{Kejun Zhang}
\authornote{Corresponding author.}
\affiliation{%
   \institution{Zhejiang University and Innovation Center of Yangtze River Delta, Zhejiang University}
  \city{Hangzhou}
  \state{Zhejiang}
  \country{China} 
}
\email{zhangkejun@zju.edu.cn}

\renewcommand{\shortauthors}{Ji et al.}

\begin{abstract}
  The burgeoning growth of video-to-music generation can be attributed to the ascendancy of multimodal generative models. However, there is a lack of literature that comprehensively combs through the work in this field. To fill this gap, this paper presents a comprehensive review of video-to-music generation using deep generative AI techniques, focusing on three key components: conditioning input construction, conditioning mechanism, and music generation frameworks. We categorize existing approaches based on their designs for each component, clarifying the roles of different strategies. Preceding this, we provide a fine-grained categorization of video and music modalities, illustrating how different categories influence the design of components within the generation pipelines. Furthermore, we summarize available multimodal datasets and evaluation metrics while highlighting ongoing challenges in the field. 
\end{abstract}





\maketitle

\section{Introduction}
Video-to-music (V2M) generation is a rapidly emerging field at the intersection of computer vision and audio synthesis, aiming to generate music that aligns semantically and temporally with video content. Studies \cite{1,2,3} have shown that a piece of melodious music can vastly attract the audience's attention and interest in watching the video \cite{4}. Traditional video soundtracks rely on professional editors to manually align video and audio or custom compositions by music producers, requiring significant resources and lacking flexibility. For video creators like vloggers, finding suitable, royalty-free background music can also be a cumbersome and time-consuming task. Additionally, while the state-of-the-art music generation tools such as Suno\footnote{https://suno.com/} \cite{5} offer high-quality outputs, their text-to-music mode struggles to achieve precise temporal alignment with video content, limiting their applicability.
\begin{figure}[t]
  \centering
  \includegraphics[width=\linewidth]{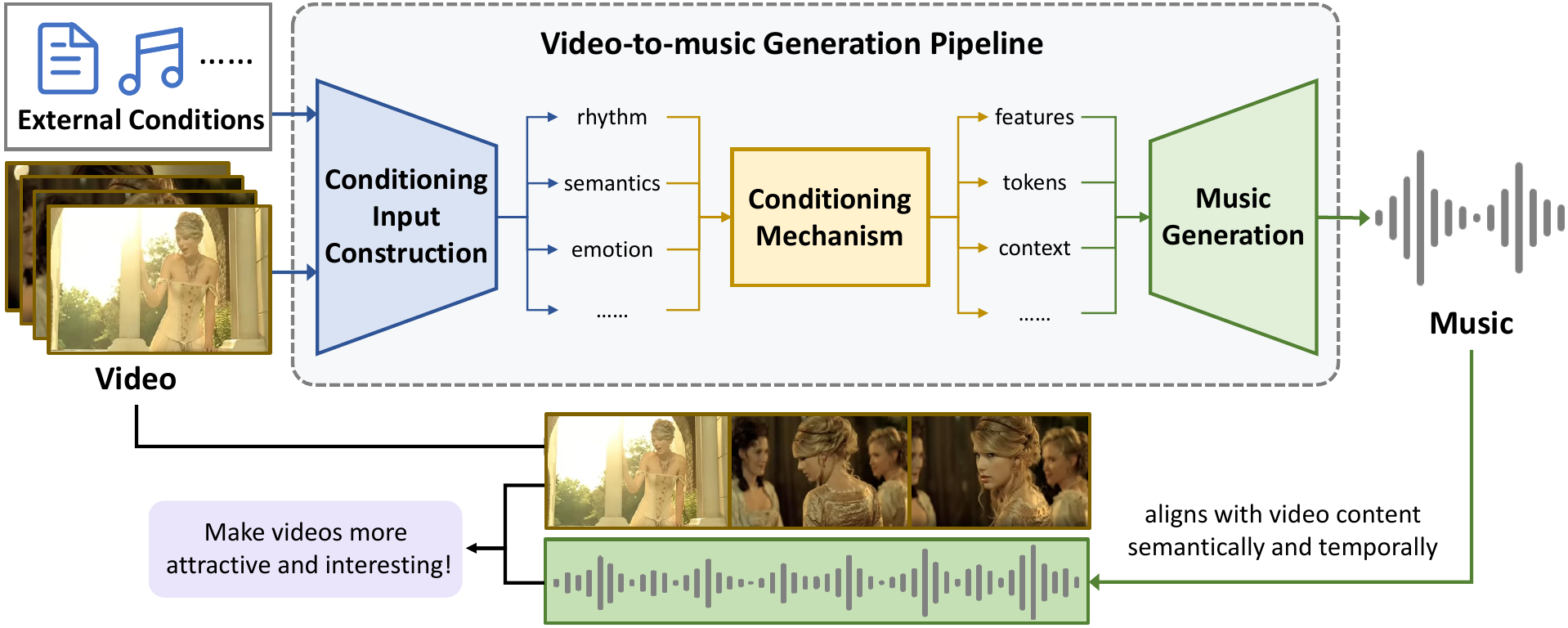}
  \caption{The overview of video-to-music generation pipeline}
  \label{fig1}
\end{figure}

Recent advancements in deep learning, especially multimodal generative models, such as multimodal diffusion models and multimodal large language models (MLLMs), have propelled progress in the realm of cross-modal generation, including V2M generation. With the growth of video streaming platforms like YouTube\footnote{https://www.youtube.com} and TikTok\footnote{https://www.tiktok.com} and the rise of automated video generation technologies (such as Sora\footnote{https://openai.com/sora/} and Veo\footnote{https://deepmind.google/technologies/veo/veo-2/}), the V2M task has gained significant attention due to its potential applications in entertainment, content creation, and personalized media experiences. For instance, automatic V2M generation can streamline the production process for filmmakers, enhance user-generated content, and enable adaptive soundtracks for interactive platforms such as virtual reality (VR) and gaming, which reduces production costs substantially. Since there are plenty of preexisting and high-quality music, some video-music retrieval (VMR) methods \cite{6,7,8} are also capable of automatically scoring music. However, it may not align well with the dynamic content of the video. Moreover, the selection is constrained by the size of the database and copyright restrictions, limiting the number of available tracks and leading to music homogenization \cite{9}.

Ensuring both the musicality of the generated content and the consistency between the music and the video makes video-to-audio generation a challenging task. The key to V2M generation lies in effectively bridging the gap between two inherently different modalities—vision and music. Videos often contain complex temporal dynamics and semantic cues, such as human movements, scene transitions, and emotional undertones, which are supposed to be translated into corresponding musical elements like rhythm, melody, and harmony. This requires sophisticated cross-modal understanding, incorporating temporal synchronization, contextual relevance, and musical creativity.

According to the prior work, the pipeline for V2M generation typically involves \textcolor{black}{three components: conditioning input construction, music generation framework, and conditioning mechanism that effectively bridges different modalities, as illustrated in Figure \ref{fig1}. As V2M generation is fundamentally a conditional generation task, conditioning input construction refers to extracting from the video the visual cues that will guide music generation, potentially complemented by additional external conditions. Conditioning mechanism describes how the model incorporates these inputs to learn the cross-modal mapping between the conditioning signals and the generated music. The music generation framework is responsible for producing the final music output and typically follows either an autoregressive or a non-autoregressive paradigm.} Note that in some studies \cite{10,11,12}, only conditioning input construction and music generation are explicitly included, even some work \cite{13,14} directly employs end-to-end models to map from video to music. The conditioning mechanisms of these studies are implicitly embedded within the models. 

Before the V2M generation gained popularity, there were already several endeavors attempting to infer plausible audio from visual signals, including tasks such as image-to-audio \cite{18}, video-to-sound effect \cite{19,20,21}, and lip-to-speech \cite{22,23,24}. Owens et al. \cite{19} synthesize plausible impact sounds from silent videos. Chen et al. \cite{18} proposed a conditional generative adversarial network to achieve cross-modal image-audio generation of musical performances. Zhou et al. \cite{20} introduced a video encoder plus a Sample RNN-based sound generator \cite{25} to generate sounds (such as baby crying, human snoring, dog) from videos in the wild, which was enhanced by REGNET proposed by Chen et al. \cite{21}. Ephrat et al. \cite{22} present an end-to-end model based on a convolutional neural network (CNN) for generating an intelligible acoustic speech signal from silent video frames of a speaking person. Kumar et al. \cite{23} propose a multi-view lipreading to audio system named Lipper. Prajwal et al. \cite{24} are the first to study lip to speech synthesis in a large vocabulary setting with thousands of words and sentences.

Albeit the similar generation pipeline, V2M generation differs from other similar cross-modal generation methods in the following aspects: 1) Compared to image-to-music, generating music that matches a video requires consideration of rhythm synchronization and temporal alignment, as videos are dynamic. 2) Unlike video-to-sound effect generation or video/lip-to-speech tasks, which demand precise synchronization of audio onsets such as matching the exact frame of a trigger pull with the corresponding gunshot sound or aligning mouth movements with spoken content, V2M generation has relatively more relaxed requirements for temporal alignment, especially for general videos. While for human-centric videos such as silent performance, there is a higher demand for precise synchronization between the human motions and the corresponding music onsets. 3) Sound effects are produced only when specific events occur, generating corresponding sounds that can abruptly change with the event, whereas music needs to maintain continuity and consistency. 4) Music generation also involves considerations of creativity and aesthetics. This includes evaluating the artistic quality and overall emotional impact of the generated music to ensure it enhances the video content effectively.

\textcolor{black}{Additionally, recent studies have explored the more advanced problem of jointly generating both video and its accompanying soundtrack, aiming to achieve multimodal co-generation and holistic consistency. For example, Ruan et al. proposed the first joint audio-video generation framework MM-Diffusion \cite{31}. Xing et al. \cite{115} proposed a multimodality latent aligner to bridge pre-trained diffusion models of single modality together to achieve audio-visual generation. You et al. \cite{111} proposed MoMu-Diffusion for long-term and synchronous motion-music generation. By contrast, the V2M generation is formulated as a conditional generation problem: the input is a pre-specified video, and the objective is to synthesize music that aligns with its semantic content, rhythmic structure, and emotional expression, thereby emphasizing the cross-modal mapping from vision to audio. Although both tasks require modeling cross-modal alignment, they differ substantially in inputs and outputs, task objectives, etc. Note that this survey focuses exclusively on V2M generation as a cross-modal music generation task, and joint video–music generation lies outside the scope of this survey.}

To the best of our knowledge, although several surveys have been conducted on unimodal music generation \cite{25,26,27,28,29}, no existing work has provided a comprehensive review specifically focused on V2M generation. \textcolor{black}{While recent surveys on multimodal music generation by Wang et al. \cite{150} and Li et al. \cite{151} also cover the V2M generation task, they only provide a coarse-grained listing of methods across different tasks, without offering a fine-grained categorization and analysis of the separate components within the V2M generation pipeline.} In contrast, this paper presents a detailed exploration of the methodologies for V2M generation\footnote{Note that in this paper, the "music" in V2M generation exclusively refers to instrumental music, excluding sound effects, speech, and other auditory elements.}, focusing on three key components: conditioning input construction, music generation, and conditioning mechanism. We categorize existing approaches based on their strategies for each component and clarify the motivations behind these choices. The components of different methods could be combined in various ways to achieve specific generation goals. This taxonomy offers researchers targeted guidance on the design of different components for V2M generation. Prior to this, we provide a fine-grained classification and explanation of the video and music in existing work, highlighting how different categories lead to the design strategies in each component. Furthermore, we summarize the datasets available in the field and commonly used evaluation metrics, finally pointing out ongoing challenges. We believe that our survey offers valuable insights for researchers in both the audio-visual field and the broader multimodal research community, particularly in exploring cross-modal relationships. From an aesthetic perspective, this survey helps explore the interplay between visual and auditory art forms and reveal the potential mechanisms of cross-sensory artistic expression. Such perspectives may inform future explorations at the intersection of technology and art, where multimodal generative models increasingly shape creative practices.

\subsection{\textcolor{black}{History and Evolution}}
\textcolor{black}{
Prior to the emergence of generative AI-driven cross-modal music generation, earlier methods primarily employed recommendation-based strategies to select soundtracks for videos from existing music libraries, including latent semantic analysis \cite{184}, heuristic ranking \cite{185}, similarity matching \cite{186}, and climax synchronization \cite{187}. While music recommendation continues to be an active area of research, concerns such as copyright issues have increasingly motivated the development of approaches that generate novel AI-composed music directly conditioned on video content.}

\textcolor{black}{
The transition of V2M methods toward generative approaches began around 2020, enabled by the emergence of relevant datasets and progress in AI-based music generation. Early studies explored automatic soundtrack generation for silent instrument performance videos \cite{10,13,14}, with outputs primarily in symbolic music formats and video types restricted to this specific domain. The release of dance dataset (e.g., AIST \cite{121}, AIST++ \cite{122}) later expanded the scope to dance videos, giving rise to a substantial body of work on dance-to-music generation \cite{46,48,54,55,72}. Since silent performances and dance both involve human body motion, which exhibit temporal synchrony with background music rhythm, prior work commonly applied pose estimation to extract body keypoints and derive rhythm-related conditioning signals for music generation. Models in this phase were largely based on convolutional neural networks (CNN), long short-term memory networks (LSTM), Transformers, and classical generative paradigms such as variational autoencoders (VAE) and generative adversarial networks (GAN).}

\textcolor{black}{
As online video platforms and video generation technologies evolved, the demand for automatic music scoring for general videos (e.g., films, vlogs, MVs) became increasingly prominent. Researchers collected larger and more diverse datasets, leading to a new wave of general video-to-music studies \cite{65,78,59,58,4,64,159,172,173}. Methodologically, the availability of powerful pretrained visual understanding models (e.g., CLIP \cite{79}, I3D \cite{56}) provided rich video features, while advancements in multimodal large models and diffusion models supplied new technical foundations for cross-modal music generation. In particular, the rapid progress of text-to-music (T2M) models (e.g., MusicGen \cite{34} and AudioLDM \cite{102}) offered strong, reusable foundation models for V2M generation. Meanwhile, commercial interest has accelerated as tools from Suno\footnote{\label{fn:sun}https://suno.com/blog/introducing-suno-scenes}, ElevenLabs\footnote{\label{fn:ele}https://elevenlabs.io/studio/video-to-music}, Adobe\footnote{\label{fn:ado}https://www.adobe.com/products/firefly/features/ai-music-generator.html}, and others now support V2M generation, reflecting the growing maturity and adoption of AI-generated music.}

\textcolor{black}{
The above brief overview of V2M research reveals that progress in this field builds not only on advances in music AI but also on broader developments in computer vision, multimodal learning, and generative AI as a whole. It also mirrors broader trends, including improvements in AI music quality, increasing market demand, and greater public acceptance of AI-generated music. Looking forward, the future V2M generation is expected to address more open-domain and contemporary challenges, enabling richer, more flexible, and more realistic multimodal generative experiences.}

\textcolor{black}{The remainder of this paper is organized as follows.
Section 2 provides a taxonomy of the two modalities involved in V2M generation, highlighting how different modality characteristics may lead to distinct component designs.
Section 3 offers a comprehensive and detailed review of existing V2M methods, structured around three core components: conditioning input construction, conditioning mechanism, and music generation, and presents a broad comparison of representative approaches from subjective, practical, and engineering-oriented perspectives.
Section 4 summarizes commonly used datasets in the V2M generation field.
Section 5 introduces widely adopted objective and subjective evaluation metrics.
Section 6 outlines some persisting challenges in V2M generation.
Section 7 discusses broader implications, including ethical considerations and real-world applications.
Finally, Section 8 concludes this survey.}

\begin{figure}[t]
  \centering
  \includegraphics[width=.7\linewidth]{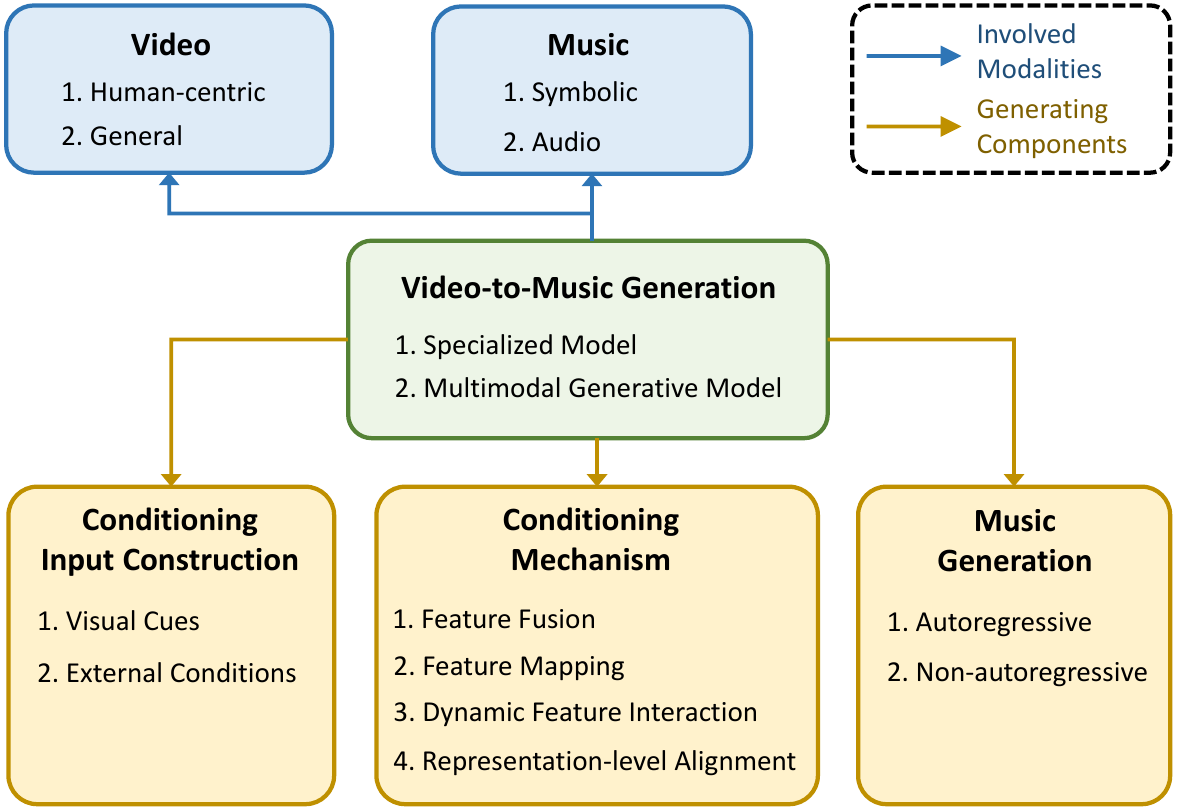}
  \caption{The involved modalities and generating components of video-to-music generation}
  \label{fig2}
\end{figure}

\section{INVOLVED MODALITY}
This section provides a fine-grained classification of the two modalities involved in video-to-music (V2M) generation, as shown in Figure \ref{fig2}. In particular, videos are usually represented by 3D signals indicating RGB values in both spatial (i.e., height $\times$ width) and temporal dimensions, while music is in 1D waveform digits or discrete symbolic events across the temporal dimension \cite{31}. The distinct classifications are related to the datasets utilized and the specific objectives targeted in various studies, influencing the design of different components in V2M generation.

\subsection{Video}
Videos can be categorized into human-centric videos and general videos based on whether they contain human movements, such as dances and sports.

\subsubsection{Human-centric.}
Human-centric videos feature rhythmic movements synchronized with the music beat, such as dance, floor exercises, and figure skating. When generating music for such videos, the primary focus is typically on whether the generated music's rhythm aligns with the movements in the video. Consequently, most studies extract rhythmic features from these videos, as detailed in Section \ref{sec:rhy}. It is worth noting that silent performance videos also fall under human-centric videos, as the instrumental playing motions also change in accordance with the musical beats.

\subsubsection{General.}
In contrast, the general videos discussed in this paper refer to those that do not explicitly include human movements synchronized with the music rhythm, such as narrative music videos (MVs), commercials, and daily life vlogs. Compared to human-centric videos, extracting rhythm from general videos is more challenging. Typically, scene transitions and salient visual changes are used to guide the generation of music rhythm, while high-level video semantics help shape the melody and emotion. Since this type of video constitutes a large portion of real-world content, recent studies have increasingly concentrated on generating background music for general videos.

\subsection{Music}
According to the representations, music can be categorized into symbolic and audio formats \cite{32}, usually derived from MIDI files and WAV files, respectively. The former represents music as a sequence of discretized events (e.g., beats and notes), while the latter processes music audio as waveforms or spectrograms. There has been extensive prior research in both the fields of symbolic music \cite{27} and audio music generation \cite{33,34,35,36,37}.

\subsubsection{Symbolic.}
Symbolic music encompasses musical events with explicit semantics, such as pitch, duration, and tempo, enabling clear correspondences with video elements like timing and motion speed, as demonstrated in approaches of CMT \cite{38}. However, symbolic music has several limitations: 1) it is difficult to collect, with datasets typically being small; 2) it lacks expressiveness and emotional depth; 3) it usually involves a limited range of instruments, restricting diversity and hindering the model's ability to synthesize complex and varied music. There are many ways to represent symbolic music, among which the most common and widely used method is MIDI event representation \cite{27}.

\subsubsection{Audio}
In contrast, musical audio is more complex and computationally expensive, lacking explicit semantic information but containing expressive and nuanced details. Additionally, it is easier to collect, as a large amount of paired video-music audio data is available online. As a result, many studies take the initial step of gathering video-music datasets from the internet to serve as training data. Due to the high dimensionality and complexity of waveforms, audio is typically first transformed into spectrograms or latent space representations before being processed by the models.

\begin{longtable}{|p{2.7cm}|l|p{1.5cm}|l|p{1.3cm}|p{1.8cm}|p{1.6cm}|p{1.7cm}|}
\caption{Methods for video-to-music generation. \colorbox{yellow!25}{FF}, \colorbox{blue!10}{FM}, \colorbox{green!10}{CA}, and \colorbox{orange!25}{RA} in the "Conditioning Mechanism" column refer to Feature Fusion, Feature Mapping, Cross Attention, and Representation-level Alignment, respectively. Table cell with \colorbox{blue!10}{background} and \colorbox{yellow!25}{background} in the "Music
Generation" column refer to Autoregressive and Non-autoregressive methods, respectively.} \label{tab1}\\
\hline
\multirow{2}{*}{Method} & \multirow{2}{*}{Year} 
& \multirow{2}{*}{Video}               
& \multirow{2}{*}{Music}   
& \multicolumn{2}{l|}{\makecell[l]{Conditioning Input}}  
& \multirow{2}{*}{\makecell[l]{Conditioning\\Mechanism}}
& \multirow{2}{*}{\makecell[l]{Music\\Generation}}  \\ \cline{5-6} 
& & &
& Visual Cues & External Conditions
& 
&    \\ \hline
\endfirsthead
\multicolumn{8}{c}%
{{\bfseries Table 1 (Continued)}} \\
\hline
\multirow{2}{*}{Method} & \multirow{2}{*}{Year} 
& \multirow{2}{*}{Video}               
& \multirow{2}{*}{Music}   
& \multicolumn{2}{l|}{Conditioning Input}  
& \multirow{2}{*}{\makecell[l]{Conditioning\\Mechanism}}
& \multirow{2}{*}{\makecell[l]{Music\\Generation}}  \\ \cline{5-6} 
& & &
& Visual Cues & External Conditions 
& 
& \\
\hline
\endhead

\hline \multicolumn{8}{r}{{Continued on next page}} \\
\endfoot

\hline
\endlastfoot
Sight to sound \cite{13} &2020 &Silent Performance &Symbolic &\cellcolor{orange!25}Spatio-temporal & &\cellcolor{blue!10}FM &\cellcolor{yellow!25}{CNN} \\ \hline
Audeo \cite{14}&2020 &Silent Performance &Symbolic &\cellcolor{orange!25}Spatio-temporal & &\cellcolor{blue!10}FM &\cellcolor{yellow!25}{CNN \& GAN}\\ \hline
Foley Music \cite{10}&2020 &Silent Performance &Symbolic &\cellcolor{yellow!25}Rhythmic & &\cellcolor{green!10}CA &\cellcolor{blue!10}{Transformer}\\ \hline
MI Net \cite{45}&2020 &silent performance &Audio &\cellcolor{yellow!25}Rhythmic & &\cellcolor{yellow!25}FF &\cellcolor{yellow!25}{VQ-VAE} \\ \hline
Dance2Music \cite{46}&2021 &Dance &Symbolic &\cellcolor{blue!10}Semantic & &\cellcolor{yellow!25}FF &\cellcolor{blue!10}{Fully Connected} \\ \hline
RhythmicNet \cite{48}&2021 &Dance &Symbolic &\cellcolor{yellow!25}Rhythmic & &\cellcolor{green!10}CA &\cellcolor{blue!10}{Transformer} \\ \hline
CMT \cite{38}&2021 &General &Symbolic &\cellcolor{yellow!25}Rhythmic &Genre, Instrument &\cellcolor{yellow!25}FF &\cellcolor{blue!10}{Transformer} \\ \hline
V-MusProd \cite{12} &2022 &\multirow[t]{2}{*}{General: MV} &Symbolic &\cellcolor{green!10}Fusion & &\cellcolor{yellow!25}FF &\cellcolor{blue!10}{Transformer} \\ 
& & & &\cellcolor{green!10} & &\cellcolor{green!10}CA &\cellcolor{blue!10}{} \\ \hline
D2M-GAN \cite{54} &2022 &Dance &Audio &\cellcolor{green!10}Fusion & &\cellcolor{blue!10}FM &\cellcolor{yellow!25}{GAN}  \\ \hline
CDCD \cite{55} &2022 &Dance &Audio &\cellcolor{green!10}Fusion & &\cellcolor{green!10}CA &\cellcolor{yellow!25}{Diffusion}  \\ 
 & & & &\cellcolor{green!10} & &\cellcolor{orange!25}RA &\cellcolor{yellow!25} \\ \hline
LORIS \cite{72} &2022 &Dance \& Sports &Audio &\cellcolor{green!10}Fusion &Genre &\cellcolor{green!10}CA &\cellcolor{yellow!25}{Diffusion}  \\ \hline
Motion2Dance \cite{75} &2023 &Dance &Audio &\cellcolor{yellow!25}Rhythmic &Genre &\cellcolor{green!10}CA &\cellcolor{yellow!25}{Diffusion}  \\ \hline
\multirow{2}{*}{\makecell[l]{MuMu-LLaMA \cite{188}\\(M\textsuperscript{2}UGen \cite{65})}} &2023 &General &Audio &\cellcolor{blue!10}Semantic &\multirow{2}{*}{} &\cellcolor{blue!10}FM &\cellcolor{blue!10}{MusicGen}\\ 
& & & &\cellcolor{blue!10} & &\cellcolor{green!10}CA &\cellcolor{yellow!25}{AudioLDM 2} \\ \hline
Dance2MIDI \cite{51} &2024 &Dance &Symbolic &\cellcolor{green!10}Fusion & &\cellcolor{green!10}CA &\cellcolor{blue!10}{Transformer} \\ \hline
DanceComposer \cite{77} &2024 &Dance &Symbolic &\cellcolor{green!10}Fusion & &\cellcolor{green!10}CA &\cellcolor{blue!10}{Transformer} \\ \hline
\multirow[t]{2}{*}{\makecell[l]{Textual Inversion \cite{17}}} &2024 &Dance &Audio &\cellcolor{yellow!25}Rhythmic &\multirow[t]{2}{*}{\makecell[l]{Dance genre}} &\cellcolor{blue!10}FM &\cellcolor{blue!10}{MusicGen} \\ 
& & & &\cellcolor{yellow!25} & &\cellcolor{green!10}CA &\cellcolor{yellow!25}{Riffusion or AudioLDM} \\ \hline
Dance2Music-Diffusion \cite{11} &2024 &Dance &Audio &\cellcolor{yellow!25}Rhythmic &Dance genre &\cellcolor{green!10}CA &\cellcolor{yellow!25}{Diffusion} \\ \hline
\multirow{2}{*}{\makecell[l]{MoMo-Diffusion \cite{111}}} &\multirow[t]{2}{*}{2024} &\multirow{2}{*}{\makecell[l]{Dance \&\\ Sports}} &\multirow[t]{2}{*}{Audio} &\multirow[t]{2}{*}{\cellcolor{yellow!25}Rhythmic} & &\cellcolor{yellow!25}FF &\multirow[t]{2}{*}{\cellcolor{yellow!25}{Diffusion}} \\ 
& & & &\cellcolor{yellow!25} & &\cellcolor{orange!25}RA &\cellcolor{yellow!25} \\ \hline
InteractiveBeat \cite{53} &2024 &Dance \& Sports \& Aerobics &Symbolic &\cellcolor{yellow!25}Rhythmic & &\cellcolor{green!10}CA &\cellcolor{blue!10}{Transformer} \\ \hline
Video2Music \cite{58} &2024 &General &Symbolic &\cellcolor{green!10}Fusion & &\cellcolor{yellow!25}FF &\cellcolor{blue!10}{Transformer} \\ \hline
Diff-BGM \cite{59} &2024 &General: MV &Audio &\cellcolor{blue!10}Semantic & &\cellcolor{green!10}CA &\cellcolor{yellow!25}{Diffusion} \\ \hline
V2Meow \cite{78} &2024 &General: MV &Audio &\cellcolor{green!10}Fusion &Text description &\cellcolor{blue!10}FM &\cellcolor{blue!10}{Transformer} \\ \hline
VidMuse \cite{4} &2024 &General &Audio &\cellcolor{green!10}Fusion & &\cellcolor{green!10}CA &\cellcolor{blue!10}{Transformer} \\ \hline
\multirow[t]{2}{*}{MuVi \cite{82}} &\multirow[t]{2}{*}{2024} &\multirow[t]{2}{*}{General} &\multirow[t]{2}{*}{Audio} &\multirow[t]{2}{*}{\cellcolor{blue!10}Semantic} & &\cellcolor{yellow!25}FF &\multirow{2}{*}{\cellcolor{yellow!25}{\makecell[l]{Flow\\matching}}} \\
& & & &\cellcolor{blue!10} & &\cellcolor{orange!25}RA &\cellcolor{yellow!25}matching \\ \hline
\multirow[t]{2}{*}{VMAS \cite{40}} &\multirow[t]{2}{*}{2024} &\multirow[t]{2}{*}{General: MV} &\multirow[t]{2}{*}{Audio} &\multirow[t]{2}{*}{\cellcolor{orange!25}Spatio-} & &\cellcolor{yellow!25}FF &\cellcolor{blue!10}{Transformer} \\ 
 & & & &\cellcolor{orange!25}temporal & &\cellcolor{orange!25}RA &\cellcolor{blue!10}\\ \hline
VidMusician \cite{15} &2024 &General &Audio &\cellcolor{green!10}Fusion & &\cellcolor{green!10}CA &\cellcolor{blue!10}{MusicGen} \\ \hline
SONIQUE \cite{16} &2024 &General &Audio &\cellcolor{blue!10}Semantic &\multirow{4}{*}{\makecell[l]{Genre,\\Instrument,\\Tempo rate,\\Melody}} &\cellcolor{blue!10}FM &\cellcolor{yellow!25}{Diffusion} \\ 
& & & &\cellcolor{blue!10} & &\cellcolor{green!10}CA &\cellcolor{yellow!25} \\ 
& & & &\cellcolor{blue!10} & &\cellcolor{green!10} &\cellcolor{yellow!25} \\ 
& & & &\cellcolor{blue!10} & &\cellcolor{green!10} &\cellcolor{yellow!25} \\ \hline
Mozart's touch \cite{70} &2024 &General &Audio &\cellcolor{blue!10}Semantic & &\cellcolor{blue!10}FM &\cellcolor{blue!10}{MusicGen} \\ 
& & & &\cellcolor{blue!10} & &\cellcolor{green!10}CA &\cellcolor{blue!10} \\ \hline
\multirow{2}{*}{\makecell[l]{Visuals Music\\Bridge (VMB) \cite{160}}} &\multirow[t]{2}{*}{2024} &\multirow[t]{2}{*}{General} &\multirow[t]{2}{*}{Audio} &\multirow[t]{2}{*}{\cellcolor{blue!10}Semantic} &\multirow{2}{*}{\makecell[l]{Retrieved ref-\\erence music}} &\cellcolor{blue!10}FM &\multirow[t]{2}{*}{\cellcolor{yellow!25}{Diffusion}} \\
& & & &\cellcolor{blue!10} & &\cellcolor{green!10}CA &\cellcolor{yellow!25} \\ \hline
PN-Diffusion \cite{152} &2025 &Dance &Audio &\cellcolor{green!10}Fusion & &\cellcolor{green!10}CA &\cellcolor{yellow!25}{Diffusion} \\ \hline
MotionComposer \cite{158} &2025 &Dance &Symbolic &\cellcolor{green!10}Fusion &Retrieved reference chord &\cellcolor{green!10}CA &\cellcolor{yellow!25}{Diffusion} \\ \hline
GACA-DiT \cite{174} &2025 &Dance &Audio &\cellcolor{green!10}Fusion & &\cellcolor{yellow!25}FF &\cellcolor{yellow!25}{Flow matching} \\ \hline
XMusic \cite{90} &2025 &General &Symbolic &\cellcolor{green!10}Fusion & &\cellcolor{yellow!25}FF &\cellcolor{blue!10}{Transformer} \\ \hline
EMSYNC \cite{163} &2025 &General: MV &Symbolic &\cellcolor{green!10}Fusion & &\cellcolor{yellow!25}FF &\cellcolor{blue!10}{Transformer} \\ \hline
GVMGen \cite{64} &2025 &General &Audio &\cellcolor{blue!10}Semantic & &\cellcolor{green!10}CA &\cellcolor{blue!10}{Transformer} \\ \hline
C\textsuperscript{3}GVS \cite{157} &2025 &General &Audio &\cellcolor{blue!10}Semantic &\multirow{2}{*}{\makecell[l]{Emotion,\\Style music}} &\cellcolor{green!10}CA &\cellcolor{yellow!25}{Diffusion} \\ 
& & & &\cellcolor{blue!10} & &\cellcolor{orange!25}RA &\cellcolor{yellow!25} \\ \hline
CoT-VTM \cite{165} &2025 &General &Audio &\cellcolor{blue!10}Semantic & &\cellcolor{blue!10}FM &\cellcolor{blue!10}{Transformer} \\ 
& & & &\cellcolor{blue!10} & &\cellcolor{green!10}CA &\cellcolor{blue!10} \\ \hline
FilmComposer \cite{155} &2025 &General: Film &Audio &\cellcolor{green!10}Fusion & &\cellcolor{yellow!25}FF &\cellcolor{blue!10}{Transformer} \\ \hline
Time-Varying Condition \cite{159} &2025 &General &Audio &\cellcolor{green!10}Fusion &\multirow{3}{*}{\makecell[l]{Beat, Melody,\\Intensity,\\Emotion}} &\cellcolor{yellow!25}FF &\cellcolor{blue!10}{Transformer} \\ 
& & & &\cellcolor{green!10} & &\cellcolor{green!10}CA &\cellcolor{blue!10} \\ \hline
\multirow{2}{*}{\makecell[l]{Decomposition and\\Alignment \cite{171}}} &2025 &General &Audio &\cellcolor{green!10}Fusion & &\cellcolor{blue!10}FM &\cellcolor{blue!10}{CNN \& CA} \\
& & & &\cellcolor{green!10} & &\cellcolor{orange!25}RA &\cellcolor{blue!10} \\ \hline
VeM \cite{172} &2025 &General &Audio &\cellcolor{green!10}Fusion & &\cellcolor{yellow!25}FF &\cellcolor{yellow!25}{Diffusion} \\ 
& & & &\cellcolor{green!10} & &\cellcolor{green!10}CA &\cellcolor{yellow!25} \\ \hline
Diff-V2M \cite{173} &2025 &General &Audio &\cellcolor{green!10}Fusion & &\cellcolor{green!10}CA &\cellcolor{yellow!25}{Diffusion} \\ \hline
DyViM \cite{170} &2025 &\multirow{2}{*}{\makecell[l]{Dance \&\\General}} &Audio &\cellcolor{green!10}Fusion & &\cellcolor{yellow!25}FF &\cellcolor{blue!10}{Transformer} \\ 
& &  & &\cellcolor{green!10} & &\cellcolor{green!10}CA &\cellcolor{blue!10} \\ \hline
\end{longtable}
\section{METHODS FOR VIDEO-TO-MUSIC GENERATION}
To date, video-to-music (V2M) generation studies have employed two types of models: specialized models and multimodal generative models, including Multimodal Large Language Models (MLLMs) and multimodal diffusion models. Specialized models are designed specifically for V2M generation. In contrast, multimodal generative models handle a broad range of multimodal tasks, making them more general-purpose but less fine-tuned for this task. As a result, specialized models typically excel in producing high-quality, domain-specific music, while multimodal generative models may sacrifice performance in V2M generation due to their broader applicability. 

Whether using specialized models or multimodal generative models, existing methods for V2M generation typically encompass three key components (as illustrated in Figure \ref{fig2}): conditioning input construction, conditioning mechanisms, and music generation. As V2M generation is fundamentally a conditional generation task, conditioning input construction refers to extracting from the video the signals that will guide music generation, potentially complemented by additional external conditions. Conditioning mechanisms describe how the model incorporates these inputs to learn the cross-modal mapping between the conditioning signals and the generated music. This section reviews existing V2M studies, focusing on these three critical components, as shown in Table \ref{tab1}. Each component can be implemented using different methods, which we have categorized into subgroups according to their goals or techniques. 

Table \ref{tab1} reveals several notable observations. First, V2M generation has gradually become a hot topic in recent years, attracting increasing attention from a growing number of studies. As methods have evolved, the fusion of visual cues that comprehensively consider multiple aspects of the video are increasingly adopted, and cross attention has emerged as a common and convenient conditioning mechanism across various music generation methods. In terms of music generation, both autoregressive and non-autoregressive modeling remain mainstream paradigms. \textcolor{black}{Moreover, external conditioning has evolved beyond static attributes (e.g., genre) to more dynamic forms, such as time-varying conditions and reference music.}

\begin{figure}[t]
  \centering
  \includegraphics[width=\linewidth]{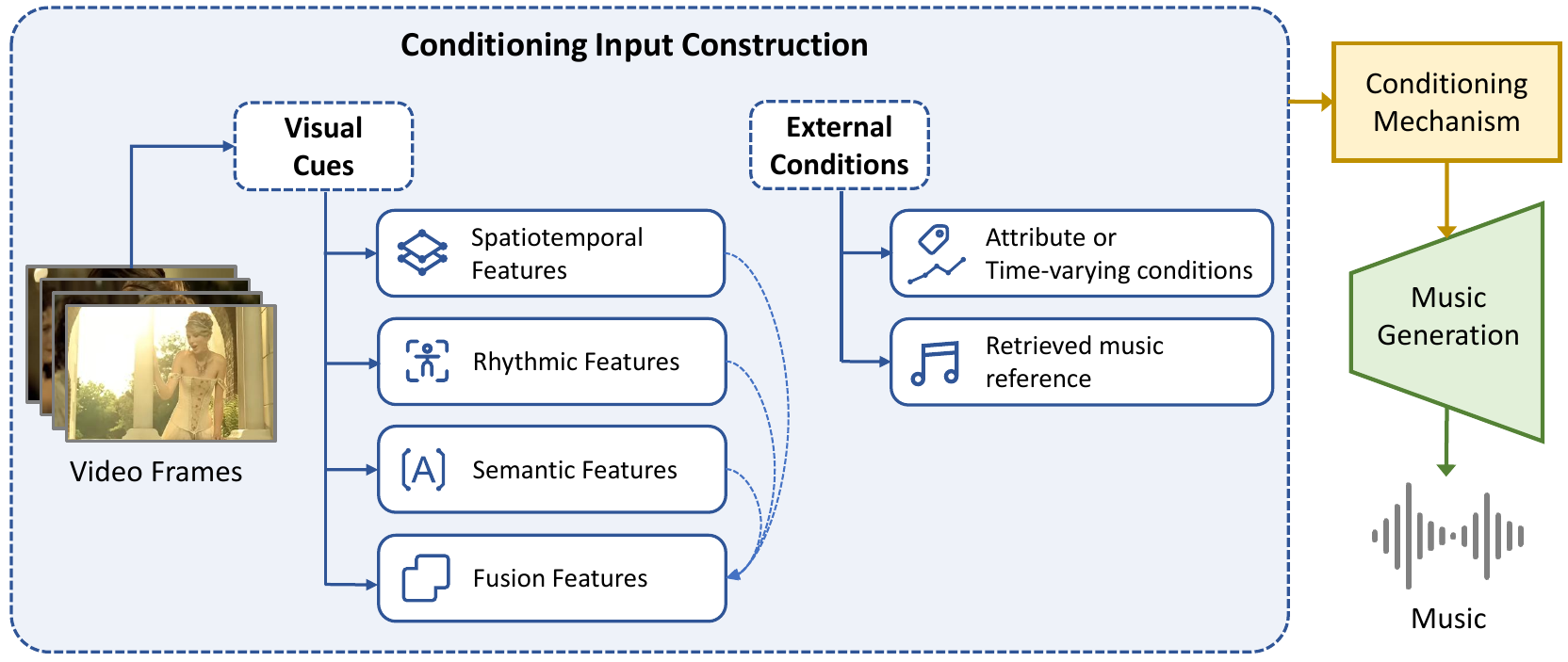}
  \caption{The taxonomy diagram of conditioning input construction}
  \label{fig3}
\end{figure}

\subsection{Conditioning Input Construction}
\textcolor{black}{In video-to-music generation, the conditioning inputs encompass both visual cues extracted from various aspects of the video, such as semantic and rhythmic features, as well as external conditions that can be specified by the user, which enhance the controllability and personalization of the generated music, as shown in Figure \ref{fig3}.}
\subsubsection{Visual Cues.}
\paragraph{Spatiotemporal Features}
To effectively capture temporal and spatial information in videos, some studies have employed convolutional networks to extract spatiotemporal features from consecutive frames. Koepke et al. \cite{13} and Su et al. \cite{14} both generated music for piano performance videos, serving as pioneering studies in the field of V2M generation. They used ResNet18 \cite{39} as the backbone to process five consecutive video frames for capturing temporal and spatial information. The key differences lie in that Koepke et al. utilized a 3D convolution-based aggregation module to integrate temporal information from consecutive frames, alongside a slope module that explicitly models the keyboard layout to preserve spatial localization. In contrast, Su et al. introduced a feature transform module to transform multi-scale features at residual blocks for detecting the visual cues on various scales better, coupled with a multi-scale feature attention network to capture spatial dependencies and semantic relevance. Additionally, Lin et al. \cite{40} modified the original Hiera model \cite{41}, a hierarchical spatio-temporal encoder that performs well on various video recognition tasks, to encode video efficiently.

\paragraph{Rhythmic Features\label{sec:rhy}}
\textcolor{black}{
Rhythmic features constitute one of the key guiding conditions in video-driven music generation, with their sources varying across different video types. In human-centric videos, rhythm is typically derived from body motion, represented via 2D keypoints \cite{43} or the 3D Skinned Multi-Person Linear model (SMPL) model \cite{42}. In general videos, rhythmic cues are usually extracted from optical flow \cite{182} or consecutive RGB frame differences, reflecting scene transitions and visual dynamics.}

\subparagraph{\textcolor{black}{\textbf{Human-centric.}}}
For synthesizing plausible music for silent instrument performance videos, Gan et al. \cite{10} first used the open-source OpenPose toolbox \cite{43} and API \cite{44} to obtain key points of the human body and hand fingers from video frames, which represent human skeleton sequence as an undirected spatial-temporal graph, where the node corresponds to a key point of the human body and edges reflect the natural connectivity of body keypoints. Afterward, the graph convolutional network (GCN) encoder is utilized to produce visual latent vectors over time. Su et al. \cite{45} used a bidirectional Gated Recurrent Unit (GRU) as the body movement encoder to encode the 2D human pose keypoints into a latent representation. For detecting musical beats, Pedersoli et al. \cite{47} extracted the skeletal body keypoints of a dancer from each video frame and use Temporal Convolutional Neural Network (TCN) architectures for beat tracking, which can guide the generation of rhythmic music.

Acknowledging the widespread use of 3D character movements in games and animations, some methods shift from encoding 2D keypoints to processing 3D motion data to broaden their application scope. For better extracting temporal information and inter-frame relationships in dance movements, Zhang et al. \cite{11} adopt the Transformer encoder to extract dance motion features from the 3D SMPL sequence. Tan et al. \cite{75} represented each frame's motion using joint positions, 6D joint orientations, and linear/angular velocities to capture both structure and dynamics. Although no explicit rhythmic cues are modeled, this rich motion encoding effectively reflects temporal variation and supports generating beat-aligned, rhythmically coherent dance music. Based on Yu et al. \cite{72}, Li et al. \cite{17} computed joint velocities from first-order differences of keypoints, discretized their directions into $K$ bins, and derived discrete accelerations by differentiating over time. Positive accelerations are summed across joints and directions to form a global acceleration curve. Local maxima on this curve are then marked as 1 to produce the final rhythm sequence.

In addition to motion features, especially for dance music generation, some studies extract rhythmic style characteristics as well. For example, Su et al. \cite{48} encoded video into rhythm correlated with human body movements and \textcolor{black}{defined "style" as transitional movements of the human body, such as rapid and sudden movements.} Specifically, keypoints are processed using a spatiotemporal GCN \cite{49} combined with a transformer encoder \cite{50} to capture motion features for estimating music beats. Given the periodic nature of music beats and the diverse visual dynamics of human movements, an additional style stream is proposed to capture fast movements. The combination of the two streams constitutes the movement rhythm. Afterward, Liu et al. \cite{53} proposed InteractiveBeat to generate evolving soundtracks in real-time for a camera input that captures the person's movements. The GRU-based VisBeatNet first predicts kinematic offsets, visual beats, and estimates tempo from a live human motion stream. Then, the adversarial style transfer module MuStyleNet converts these kinematic offsets into a drum "style" vector representing the rhythm of the generated drum. Compared to Su et al. \cite{48}, this method is more effective in bridging the motion rhythm and drum rhythm and works in real-time.

\subparagraph{\textcolor{black}{\textbf{General.}}}
Although motion features support rhythmically coherent music generation for human-centric videos, they are not applicable to general videos lacking human movements, limiting their applicability to specific scenarios and hindering their effectiveness for general videos with diverse content \cite{58}. To address this, alternative approaches have been developed to derive rhythm features suitable for a wider array of video types. Di et al. \cite{38} addressed the task of video background music generation without relying on paired video and music data. Specifically, they established three rhythmic relationships between video and music, i.e., video timing and music beat, motion speed and simu-note density, motion saliency and simu-note strength. Rhythmic features are then extracted from MIDI music during training or from the video during inference. This method can generate music from arbitrary video clips. However, it ignores the semantic-level correspondence, which sometimes results in conflicting styles \cite{12}.

\paragraph{Semantic Features.}
\textcolor{black}{
Although rhythmic features provide fine-grained temporal guidance, they overlook the global semantics conveyed by the video, such as atmosphere and emotion. These semantic cues are essential for general videos where strict beat alignment is less important. Even for human-centric videos, extracting semantic features is also helpful for capturing high-level movement characteristics.
}

\subparagraph{\textcolor{black}{\textbf{Human-centric.}}}
Aggarwal et al. \cite{46} computed a dance similarity matrix from the extracted pose sequence, which is processed by CNN layers and global average pooling to get the dance feature representation that captures the semantic structure of body movements. \textcolor{black}{In addition, the I3D model \cite{56} pre-trained on Kinectics \cite{57} for video action recognition is frequently used to extract motion-related semantic features, as in \cite{54,55,72,152,158,174}.}

\subparagraph{\textcolor{black}{\textbf{General.}}}
With advances in visual understanding, more and more pre-trained models have been leveraged to extract semantic features from videos, enabling a richer understanding beyond motion and rhythm. These models, often trained on large-scale datasets, capture high-level concepts and contextual information, allowing researchers to focus on cross-modal alignment between the extracted features with music rather than extensive domain-specific training. Li et al. \cite{59} extracted visual features from each frame using a pre-trained Video CLIP \cite{60} and language features from video captions using a pre-trained BERT encoder \cite{61}. Li et al. \cite{82} explored multiple visual extractors, such as CLIP \cite{79}, VideoMAE \cite{83}, VideoMAE V2 \cite{84}, CAVP \cite{85}, and proposed a visual adaptor to compress the video features to align with the frame length of the audio features. Su et al. \cite{78} also explored various visual features from pure visual, multimodal, and quantized models, corresponding to I3D \cite{56}, CLIP \cite{79}, and ViT-VQGAN \cite{80}, respectively. These visual features can be combined to guide music generation. Zuo et al. \cite{64} used a pretrained ViT-L \cite{66} with spatial self-attention to extract deep visual features. \textcolor{black}{Qi et al. \cite{157} mapped video features into the music and sound-effect feature spaces using two MLPs trained with contrastive learning. After pretraining, these networks produce visual-based control features for music and sound effects, which are flattened and used as separate conditioning signals. As a foundational visual–language representation model, CLIP has been widely adopted \cite{159,165,167} for extracting semantic features. }

Multimodal models capable of V2M generation also utilize pre-trained models for semantic feature extraction. Zhang et al. \cite{16} extracted video description from input video using Video-LLaMA \cite{62}, which is then passed through an LLM to generate simple, descriptive musical tags. By leveraging descriptive tags, this method avoids the use of paired video-music datasets. The MuMu-LLaMA (or M\textsuperscript{2}UGen) proposed by Liu et al. \cite{188,65} integrated LLM's abilities to comprehend and generate music for different modalities. They use frozen ViT \cite{66}, ViViT \cite{67}, and MERT \cite{68} to encode the image, video and music modalities, respectively and develop multi-modal understanding adapters to align these embeddings with LLaMA 2 \cite{69}. Li et al. \cite{70} proposed Mozart's touch, a multi-modal music generation framework capable of generating music aligned with cross-modal inputs such as images, videos, and text. They use BLIP \cite{71} to generate video-level captions, which are fed into an LLM to construct a musical descriptive prompt. \textcolor{black}{Wang et al. \cite{160} converted videos into detailed music descriptions with a multimodal music description model built upon InternVL2 \cite{161} to guide music generation.} Although these approaches are capable of achieving semantic alignment, they do not guarantee temporal rhythm synchronization due to the lack of rhythm-related feature extraction from the videos.
\paragraph{Fusion Features.}
Focusing solely on rhythmic alignment while neglecting semantic correspondence may lead to conflicting styles, thereby diminishing immersion and experience. Conversely, prioritizing semantic matching while ignoring rhythmic synchronization may result in temporal mismatches between music and video actions or scene transitions, causing unnatural coherence and disrupted perceptual flow. To achieve multi-faceted consistency in V2M generation, it is essential to balance rhythm and semantic alignment. Consequently, recent studies extract both rhythmic and semantic features and fuse them to enhance overall alignment.

\textcolor{black}{Zhu et al. \cite{54,55} used a convolution-based motion encoder to encode body motions and extracted the I3D semantic features \cite{56}. The motion and I3D features are concatenated to form the final continuous conditioning input. Sun et al. \cite{152} also concatenated motion information extracted by ST-GCN with I3D visual embeddings for dance-to-music generation.} Extending from dance to multiple sports scenarios such as floor exercise and figure skating, Yu et al. \cite{72} introduced a series of context-aware conditioning, including human motions, video frame, and genre. They extracted visual rhythms from cadence movements, applied a Hawkes Process \cite{73,74} to capture temporal context, and modeled temporal relationships via a Bi-LSTM over I3D features. \textcolor{black}{Wang et al. \cite{158} predicted music beats by detecting visual rhythm, which is a binary vector computed by utilizing directogram to represent motion changes, extracting the impact envelope and picking peaks, while visual embeddings were extracted using the I3D model. Liu et al. \cite{170} extracted frame-wise dynamics features by comparing neighboring frames to compute correlation features and flow features, fusing them into motion features, and then using self-attention and pooling to obtain frame-wise dynamics features relevant to music, while high-level semantics are extracted using CLIP. Wang et al. \cite{174} extracted the I3D semantic features and proposed a Genre-Adaptive Rhythm Extraction (GARE) module to extract fine-grained dance rhythm by computing joint-level motion, applying multi-scale wavelet analysis, weighting rhythm-informative joints per genre, and fusing temporal–spatial motion features through attention.}

In some studies, global features are often extracted as semantic information, while local features serve to capture rhythmic patterns. Tian et al. \cite{4} first encoded video frames with CLIP, then used a Long-Short-Term Visual Module to learn spatial-temporal features, where the long-term module models the entire video to capture the global context while the short-term module learns the fine-grained clues at the clip level. Then, an LST-Fusion integrates long-term and short-term features via cross-attention, where short-term features query global information. Li et al. \cite{15} also extracted global and local features as high-level semantic features and low-level rhythmic features. The semantic features are derived from the CLS token of CLIP, while the average inter-frame similarity of local features extracted by CLIP is taken as the rhythm condition. \textcolor{black}{Wu et al. \cite{159} extracted three distinct video features: patch-level fine-grained image features, frame-level visual features via CLIP, and context-aware visual features via VideoMAE V2 \cite{84}, which are fused through a combination of self-attention, cross-attention, and learnable weighting, preserving fine-grained details while integrating broader contextual information. Tong et al. \cite{172} adopted a hierarchical video parsing strategy at the global, storyboard, and frame levels to extract semantic, structural, and rhythmic cues. Global captions and emotion tags provide high-level semantics. Storyboard segmentation offers local descriptions, timestamps, and durations. Frame-level transition detection supplies fine-grained rhythmic boundaries. All parsed information is encoded using CLAP for textual semantics, MAViL \cite{190} for visual content, a continuous-time MLP for temporal attributes, and binary indicators for scene transitions, forming a unified multimodal condition for music generation.}

Beyond semantic and rhythmic features, videos can also provide style-related cues. Han et al. \cite{51} utilized two branches to extract dance movement and style features, which are concatenated to form conditional control information. The video is first processed with Mediapipe \cite{52} to obtain joint coordinates, which are then encoded via a GCN and a GCN-GRU network to capture spatiotemporal movement and style embeddings, respectively. Liang et al. \cite{77} proposed DanceComposer to achieve rhythmic alignment and stylistic matching simultaneously with a Music Beats Prediction Network (MBPN) and a Shared Style Module (SSM), respectively. Specifically, skeleton sequences extracted by OpenPose are processed by a spatial-temporal GCN (ST-GCN)  and a Transformer encoder to predict music beats. The SSM encodes keypoints and log mel-spectrograms into a unified cross-modal style space, preventing similar melodies across different dances. The predicted beats and cross-modal style features jointly guide the music generation. 

To provide explicit emotional guidance for generated music, emotion-related features are extracted from video content. Zhuo et al. \cite{12} extracted semantic, color, and motion features, with semantic and color features capturing the video's theme and emotion, and motion features guiding rhythmic control. Specifically, semantic features are encoded using pretrained CLIP2Video \cite{87}, color features are represented via the 2D color histogram of each frame \cite{88}, and motion features are computed from RGB differences to determine tempo, followed by Tempo Embedding and Timing Encoding \cite{38}. Kang et al. \cite{58} also extracted semantic, scene offset, motion, and emotion features from each frame of music videos. Concretely, the semantic features are obtained using the CLIP, which also provides the probability distribution of different emotion classes (`exciting,' `fearful,' `tense,' `sad,' `relaxing,' and `neutral'). The PySceneDetect library \cite{89} is utilized for accurately detecting shot changes in videos, and scene offsets are calculated based on the detected scene IDs. Motion features are derived from the RGB difference between consecutive frames, averaged over all pixels. Tian et al. \cite{90} proposed XProjector to parse prompts of various modalities into symbolic music elements within the projection space to generate matching music. By formulating heuristic rules, XProjector maps scene transitions into tempo, optical-flow intensity into note density, and visual beat saliency into beat strength. The video's emotional category is obtained through sentiment analysis. \textcolor{black}{Sulun et al. \cite{163} used a pretrained emotion classifier to extract emotion probability distributions from videos and mapped them to the valence–arousal plane using a Gaussian mixture model. Each emotion category is modeled as a Gaussian with predefined mean and variance, and the final valence–arousal values are derived by sampling from the mixture or taking its weighted mean. Besides emotions, they introduce scene boundaries extracted using ffmpeg\footnote{ https://www.ffmpeg.org/} for temporal conditioning. To achieve finer granularity in controllable V2M generation, You et al. \cite{171} segmented video frames with SAM \cite{179} and extracted multiple features, such as semantic features, area, start position, movement, and color. For silent film clips, Xie et al. \cite{155} used ChatGPT-4V \cite{189} to recognize visual attributes (setting, brightness, color hue, action, emotional tone, view scale, and theme) and open-source algorithms to extract motion speed, motion saliency, shot cut, and plot development. Besides, they devised a controllable rhythm transformer to predict rhythm points from videos. Moreover, Ji et al. \cite{173} integrated the CLIP-based semantic features, color histogram-based emotional features, and the predicted explicit musical rhythm (including low-resolution mel-spectrogram, tempogram, and ODF) as the conditions for V2M generation.}

\subsubsection{\textcolor{black}{External Conditions}}

\textcolor{black}{
The relationship between video and music is inherently many-to-many: pairing the same video with music of different tempos or emotions may substantially alter its overall interpretation. Consequently, video-driven cues alone may be insufficient for determining appropriate music. To enable finer control, external conditions can be specified, including explicit musical attributes such as genre, emotion, tempo, and intensity, as well as implicit guidance in the form of reference examples. Note that these attributes here can be externally specified, in contrast to the aforementioned approaches that derived similar information (e.g., style, emotion) directly from the video.}

\textcolor{black}{
As a crucial musical attribute, genre has been incorporated as an external condition to guide music generation, together with video-derived features. For example, Di et al. \cite{38} took the genre and instrument type as initial tokens of CMT model to control the global music properties, which can be specified by users. In LORIS \cite{72}, one-hot genre labels are embedded and then fed into the model via cross attention. Tan et al. \cite{75} also conditioned the model with music genre as a high-level control, which is concatenated with the motion features as the conditioning signal. In dance-to-music generation, the dance genre is also used as an external condition. Li et al. \cite{17} mapped one-hot dance genre vectors into the textual space, while Dance2Music-Diffusion \cite{11} used genre as a supervisory signal for motion encoder training. V2Meow \cite{78} allows users to optionally provide a music-related text description to impose on personal preference. The music-text joint embedding model MuLan \cite{81} is used to extract audio embedding during training and text embedding during inference. Qi et al. \cite{157} imposed emotion and style conditions via score-guided noise iterative optimization during the inference stage. Wu et al. \cite{159} developed the first V2M generation model with multiple time-varying conditions: beat, melody, intensity, and emotion, offering fine-grained control over V2M generation. These control signals can be obtained via extracting using models or annotated by a music creator. Additionally, to provide reference examples for music generation, retrieval-augmented generation (RAG) techniques have been leveraged to select relevant music pieces. Wang et al. \cite{160} designed a dual-track music retrieval framework that computed similarities using CLIP or CLAP embeddings for broad retrieval and conducted a targeted retrieval based on musical attributes. To enhance stylistic chord generation, Wang et al. \cite{158} developed RAGate to dynamically retrieve the most relevant examples based on the I3D visual embedding. The retrieved music features, together with other conditioning inputs, jointly guide the music generation process.
}

\subsubsection{\textcolor{black}{Summary}}

\textcolor{black}{
Based on the above review, the extraction of video cues for V2M generation has gradually evolved from early spatiotemporal features to the combinations of multi-perspective features, e.g., semantic and rhythmic features. This evolution has been driven by two main factors. First, the scope of video types has expanded from silent performance and dance videos, i.e., human-centric videos, to a broader range of general videos. Second, the rapid development of visual understanding models has substantially improved the accessibility and quality of high-level visual features. }

\textcolor{black}{
For early research focused on human-centric videos, such as dance or silent instrument performances, extracting spatiotemporal features or rhythm-related features alone was sufficient to guide music generation that aligned with the video content. As video types expanded to general videos, however, the need to capture multi-view video information became apparent, particularly semantic and rhythmic features. These two types of features exhibit a clear trade-off: semantic features excel at capturing the global atmosphere and mood of a video but may overlook fine-grained dynamics related to rhythm. Conversely, rhythmic features are critical in dance and other human-centric scenarios but are less expressive in static or semantically oriented videos. Together, they provide complementary information that enriches video understanding, while other video-driven factors such as style and emotion further enhance the conditioning. }

\textcolor{black}{
Different video types call for different feature extraction strategies. For rhythmic features, human-centric videos typically rely on pose estimation methods such as OpenPose or MMPose \cite{17} to extract keypoints, which are then encoded using models like GCNs or TCNs. As for general videos, rhythmic cues can be derived from optical flow or frame differences and further processed using rule-based or network-based methods. For semantic features, existing methods almost leverage pretrained video understanding models. For instance, I3D is widely adopted for human-centric videos, whereas CLIP is frequently used for general videos. However, relying solely on video-driven cues is often insufficient to determine suitable music and lacks control over generation results. To address this, external conditions such as musical attributes and reference music are introduced to enhance controllability and diversity.
}
\begin{figure}[t]
  \centering
  \includegraphics[width=\linewidth]{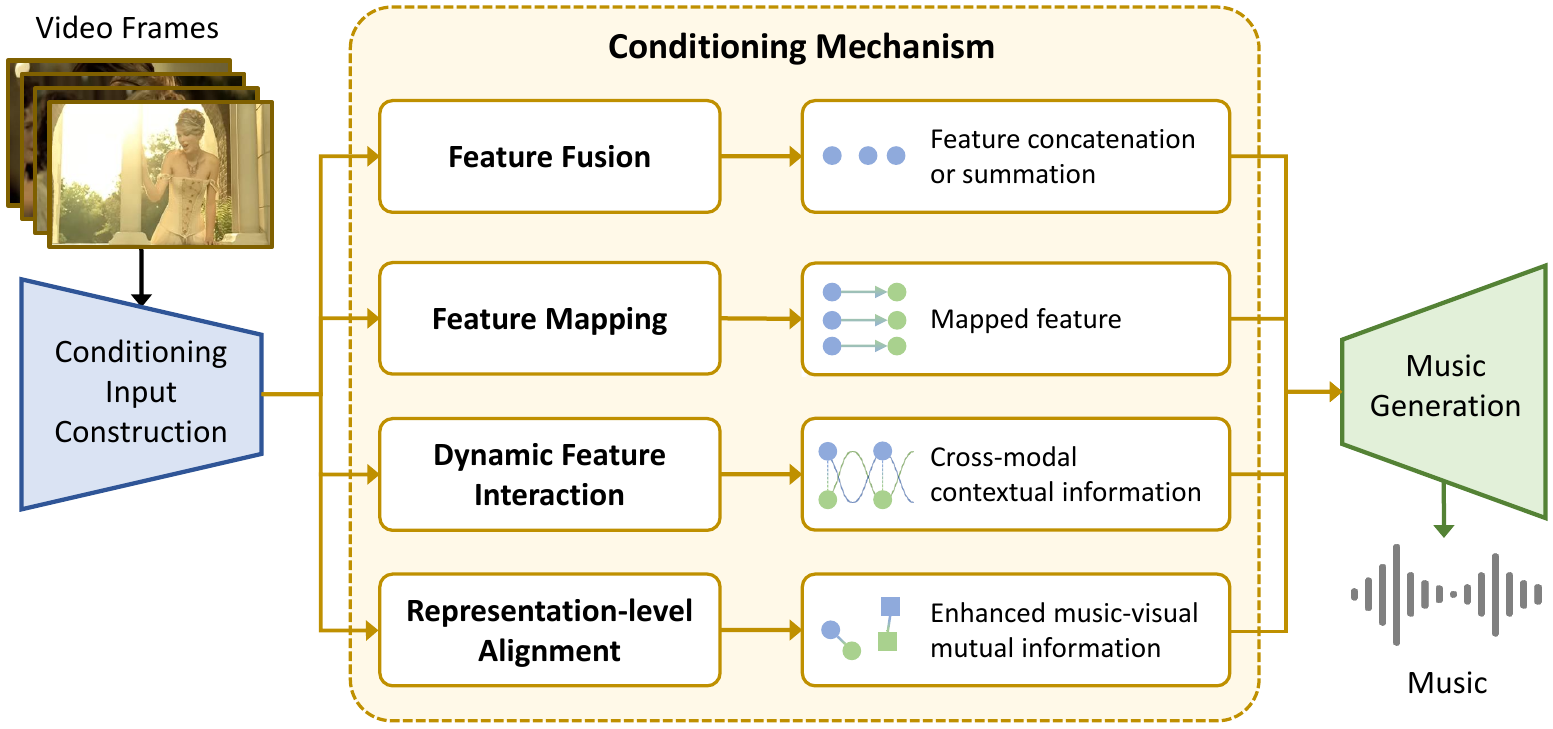}
  \caption{The taxonomy diagram of conditioning mechanism}
  \label{fig4}
\end{figure}
\subsection{Conditioning Mechanism}
To achieve video-to-music generation, a crucial component is required to bridge these two modalities, which is referred to as the conditioning mechanism in this survey. This component establishes a correspondence between video and music, enabling the model to produce music that is both semantically consistent and rhythmically synchronized with the input video. As shown in Figure \ref{fig4}, \textcolor{black}{from the perspective of cross-modal alignment learning, this survey divides conditioning mechanisms in V2M generation into four categories:
\begin{itemize}
\item[(1)] Feature Fusion, where visual features are integrated into the music generator via simple operations or modulation mechanisms, such as summation, concatenation, and Adaptive LayerNorm (AdaLN), allowing the network to implicitly learn cross-modal correspondences;
\item[(2)] Feature Mapping, where visual features are explicitly projected into the musical space (acoustic or symbolic domain) or into a textual space that serves as an intermediate conditioning bridge for music generation.
\item[(3)] Dynamic Feature Interaction, where the model employs a cross-modal attention mechanism that explicitly learns correspondences between video and the music, enabling dynamic information integration;
\item[(4)] Representation-Level Alignment, where video and music are aligned at the representation or feature level, predominantly through similarity-based objectives.
\end{itemize}
}
\subsubsection{Feature Fusion.}

\textcolor{black}{This section introduces the most straightforward conditioning mechanism, which involves fusing video features into the music generator either through simple operations, such as summation or concatenation, or via modulation mechanisms, such as Adaptive LayerNorm (AdaLN). In all cases, the fused features are directly consumed by the music generation network without explicitly modeling cross-modal interactions.}

For feature concatenation, Su et al. \cite{45} concatenated the last hidden state of the body keypoints encoder to every time step of the VQ-VAE discrete latent representation and fed the concatenated features into the Transformer encoder for predicting autoregressively. Aggarwal et al. \cite{46} concatenated the encoded dance feature with note sequences up to time $t$ as input to predict the note for time $t+1$. You et al. \cite{111} concatenated the noisy input with the embedded conditional inputs and the embedded diffusion timesteps along the temporal dimension. The fused input is then padded to match a specified maximum length and combined with positional embeddings prior to being processed by the DiT model. \textcolor{black}{Xie et al. \cite{155} sequentially prepended the processed rhythm and text conditions to the Transformer input. Wu et al. \cite{159} positioned the semantic features before the first music token, ensuring that the subsequent generation process is conditioned on a holistic video representation. Sulun et al. \cite{163} concatenated the input token sequence with emotion inputs along the sequence dimension and concatenated the learned positional encoding with boundary offsets along the feature dimension. The concatenated results are summed up as the transformer's input. Wang et al. \cite{174} concatenated the noised latents with the temporally aligned rhythm features, I3D video features and time step as the input of the DiT-based V2M framework.} Other approaches, such as CMT \cite{38}, embed extracted rhythmic features into the music representation to generate rhythmically aligned outputs, while XMusic \cite{90} places visual condition tokens at specified positions within symbolic music sequences to control generation.

Element-wise addition offers another approach to fuse conditioning inputs. In video-to-symbolic music generation \cite{12,38}, the time embeddings extracted from video are typically added to the token embeddings and positional encodings to guide the temporal structure of music generation. Li et al. \cite{82} transformed the visual condition and music latent representation into the same dimension and summed element-wise as the input of the DiT generator \cite{116}. This element-wise summation facilitates point-to-point precise alignment. \textcolor{black}{Liu et al. \cite{170} extended the existing music codes derived from RVQ codebooks with interpolated dynamic features and summed them up to obtain the music codes extension for MusicGen. As a notable exception, Tong et al. \cite{172} aligned video and music rhythms by detecting visual transitions and music beats, taking their intersection as aligned timestamps, extracting corresponding rhythmic features, and injecting them into the music latent space via AdaLN-based adaptive normalization for precise cross-modal rhythm synchronization. }

\subsubsection{\textcolor{black}{Feature Mapping.}}
\textcolor{black}{
Feature mapping methods project video inputs into another modality, either by directly mapping video to music representation or by first translating it into intermediate text-bridging or LLM-compatible embeddings, which then serve as the condition for text-to-music generation.}

\textcolor{black}{
Several studies have directly transformed visual features into music representations. For example, Koepke et al. \cite{13} and Su et al. \cite{14} fed video information into an end-to-end network that outputs music (midi or piano roll). Zhu et al. \cite{54} input the fused motion-visual features derived from dance videos into the VQ generator of GAN to predict the audio VQ representations. Su et al. \cite{78} adopted the AudioLM \cite{33} pipeline, thereby learning the mapping from visual inputs to the music semantic tokens first, which are then mapped to coarse and fine acoustic tokens. You et al. \cite{171} mapped the video features to music features by first applying dense layers and temporal self-attention, then predicting pitch, loudness, chroma, and timbre with a non-autoregressive Transposed-Conv1D network. These music features are taken as conditioning inputs for music generation.}

\textcolor{black}{Other works use text as an intermediate modality, mapping visual conditions into textual space and then leveraging pretrained text-to-music foundation models. Inspired by the textual inversion in computer vision \cite{66}, Li et al. \cite{17} developed a dual-path rhythm-genre inversion that encodes the rhythm and genre of a dance motion sequence into learnable pseudo-words within a fixed prompt template, adapting to varying rhythms and genres for personalized music generation. Zhang et al. \cite{16} mapped the video description extracted by Video-LLaMA \cite{62} into descriptive musical tags via an LLM. Then CLAP \cite{63} processes these tags and any additional user-provided tags (e.g., instruments, genres, tempo rates, and melodies) for customization. Similarly, Li et al. \cite{70} aggregated frame-level captions from BLIP \cite{71} into a video-level caption, which is then processed by the LLM to form a music-descriptive prompt. Wang et al. \cite{160} also converted videos into detailed music descriptions with a multimodal music description model. }

\textcolor{black}{
In addition, another research uses LLM as a bridging layer to connect video representations with music generation. In MuMu-LLaMA \cite{188}, video data is first encoded into spatio-temporal embeddings and then integrated into an LLM-compatible representation by a multi-modal understanding adapter. This adapted video representation is fed into LLaMA 2 for cross-modal contextual interpretation, and the hidden embeddings of the output audio tokens are transformed by an output projection layer into conditioning signals for downstream music decoders like MusicGen or AudioLDM 2.
} 

\subsubsection{Dynamic Feature Interaction}
In this survey, dynamic feature interaction refers to the mechanism by which one modality dynamically influences another through cross attention. Cross attention is first introduced as part of the Transformer architecture, enabling the decoder to focus on the encoder's outputs when generating predictions. It is now widely used in conditional generation tasks to integrate contextual information from one modality into another. Especially in V2M generation tasks, cross-attention is commonly used as a conditioning mechanism in the following two scenarios. 

First, when utilizing Transformer-based architectures as generators, the conditions are usually fed into the decoder's cross-attention module. For example, Gan et al. \cite{10} input extracted visual features into the cross-attention module of a decoder-only Transformer designed for decoding MIDI. Su et al. \cite{48} fed encoded drum tracks into a Transformer-XL decoder to generate piano or guitar instrumentation. Han et al. \cite{51} fed the dance condition into the cross-attention module of the transformer for generating drum notes. Conditioned on the style features obtained by fusing color features and semantics features via cross attention, Zhuo et al. \cite{12} used a decoder-only transformer to generate chords first. Then, adopting the same conditioning mechanism, they generated melodies conditioned on chords and finally generate accompaniments conditioned on chords and melodies. Kang et al. \cite{58} fed the video embedding into the transformer encoder, which serves as the conditioning for the transformer decoder to generate chord sequences. Tian et al. \cite{4} and Lin et al. \cite{40} conditioned autoregressive decoders on video segments using cross-attention. Li et al. \cite{15} adopted a two-stage V2M training strategy that progressively integrates global and local video features into the generator. High-level semantic features are injected through cross-attention, while low-level rhythmic features are incorporated via in-attention using zero and identity initialization. Liang et al. \cite{77} conditioned the Drum Transformer and Multi-track Transformer on music beats and the fused embedding sequence of the drum track and style feature, respectively, via cross attention. Zuo et al. \cite{64} transformed hidden visual features into musical features through both spatial and temporal cross-attention.

The second scenario involves diffusion-based generators, where visual features are typically fed as conditioning inputs into the cross-attention layers of the denoising network. For example, Tan et al. \cite{75} concatenated motion features and genre labels to guide the denoising network through cross-attention layers. Zhu et al. \cite{55} fed the motion and visual conditioning input into the cross-attention module of a transformer-based contrastive diffusion model. Zhang et al. \cite{11} fused the encoded dance motion and genre features and fed them into the cross-attention of the latent dance-to-music diffusion generator. In LORIS \cite{72}, visual, motion, and genre features are conditioned via a hierarchical cross-modal attention block. Li et al. \cite{59} found that during the unconditional music generation process, models tend to generate the melody first, followed by the rhythm. Most existing models, however, lack interpretable control signals reflecting this process. To address this, they introduced a feature selector to apply different conditions at different timesteps, and used segment-aware cross-attention with a masking mechanism to align music rhythm with video in a fine-grained, short-term manner. \textcolor{black}{Sun et al. \cite{152} input both positive conditioning and negative conditioning into the cross attention within the adapted Unet of stable diffusion \cite{153}. Tong et al. \cite{172} proposed storyboard-guided cross-attention (SG-CAtt) to preserve semantic alignment and temporal synchronization by conditioning diffusion queries on concatenated global and storyboard-level features and masking attention within storyboard boundaries. Ji et al. \cite{173} designed a hierarchical cross-attention module to effectively integrate multiple features. Specifically, emotional features are first integrated through a cross-attention layer to guide the overall mood of the generated music. Then, the semantic and rhythmic features are processed independently via parallel cross-attention and adaptively fused using a set of timestep-aware fusion strategies including Feature-wise Linear Modulation (FiLM) and weighted fusion. Many other studies \cite{157,158,159,160,167,168,170} similarly guide music generation by injecting conditioning signals via cross-attention, and are not detailed here for brevity.}

\subsubsection{Representation-Level Alignment}
Representation-level alignment methods aim to enforce correspondence between video and music at the representation or feature level, ensuring that the generated music closely follows the temporal and semantic patterns of the input video. 

Contrastive learning has been employed to enhance alignment and synchronization across different modalities. Zhu et al. \cite{55} explicitly enhanced input-output connections by maximizing their mutual information using a Conditional Discrete Contrastive Diffusion (CDCD) loss. They introduce step-wise parallel diffusion with intra-negative samples to capture intra-sample cues (e.g., rhythm), and sample-wise auxiliary diffusion with inter-negative samples to enhance instance-level distinctions (e.g., genre), achieving much faster convergence and inference than standard DPMs. Li et al. \cite{82} proposed a contrastive music-visual pretraining strategy aimed at more precise synchronization. In addition to using music and visual features from different videos as negative samples, two new sets of negative pairs are proposed: one is created by temporally shifting the original music waveform to obtain shifted music features, and the other replaces the original clip's waveform with another clip from the same music. Lin et al. \cite{40} incorporated two strategies to align video and music. The first is a global video-music contrastive objective, which guides the music to correspond to high-level video cues, such as genre. The second is a fine-grained video-beat alignment scheme to synchronize music beats with low-level video cues, such as scene transitions and dynamic human movements, where the overlap vector between the music and video beats serves as the importance weights to be incorporated into the autoregressive generation objective. You et al. \cite{111} claimed that the random selection process for constructing negative samples risks capturing similar rhythmic sequences. To address this, they proposed rhythmic contrastive learning, which constructs contrast pairs based on a kinematic amplitude indicator derived from spatial motion directrogram differences as detailed in \cite{118}, which quantifies the temporal variation in motions. Motion-music clips are categorized by their maximum kinematic amplitude, and negative samples are randomly drawn from different categories to ensure effective temporal and rhythmic alignment.

\textcolor{black}{Moreover, Qi et al. \cite{157} achieved rhythm alignment by extracting two temporal curves from music and video and matching them. Music rhythm is represented by frame-wise quantized pitch values obtained after continuous wavelet transforms, while video rhythm is represented by frame differences computed between adjacent frames. The cosine distance between the pitch curve and the video-variation curve serves as a guidance score in the score-based inference optimization to encourage synchronization. You et al. \cite{171} introduced a flow-matching based alignment (FMA) method to ensure temporal and semantic consistency. FMA treats video features as conditioning signals and models the evolution of music embeddings through a learned velocity field. By training a velocity model to approximate the ideal cross-modal flow, they measure whether the predicted music trajectory follows the video-conditioned flow, yielding an alignment loss that enforces video–music correspondence. }

One important point to emphasize is that the four conditioning mechanisms introduced above are not mutually exclusive. In fact, multiple mechanisms can be integrated to incorporate multiple conditions into the model, as exemplified in the aforementioned studies \cite{12,16,17,40,55,70,82,111,157,159,160,165,170,171,172,188}.
\begin{figure}[t]
  \centering
  \includegraphics[width=\linewidth]{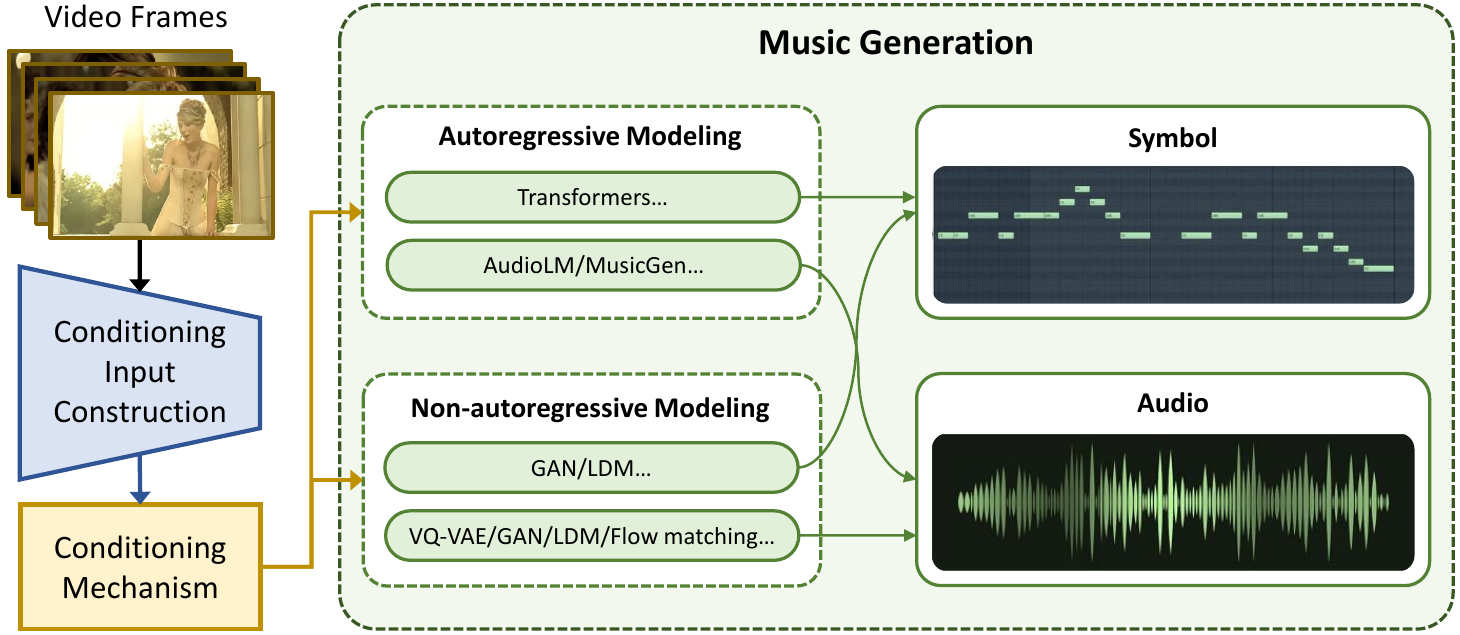}
  \caption{The taxonomy diagram of music generation}
  \label{fig5}
\end{figure}
\subsection{Music Generation}
Music generation methods can be broadly categorized into two types based on their generative paradigms: autoregressive and non-autoregressive modeling, as depicted in Figure \ref{fig5}. Autoregressive methods are often applied to predict MIDI music events or discrete audio tokens, while non-autoregressive methods excel at handling image-like audio spectrograms or latent features derived from encoding. As one of the most prevalent generative models nowadays, diffusion models \cite{92} are commonly used in non-autoregressive approaches.

\subsubsection{Autoregressive Modeling.}

\paragraph{Symbolic.}
Transformers are widely used in the symbolic music generation for given videos. Given the extracted motion features, Gan et al. \cite{10} used a decoder-only transformer \cite{93} to autoregressively predict MIDI events proposed by Oore et al. \cite{94} and then utilized a standard synthesizer (i.e., FluidSynth) to generate the music waveforms. Su et al. \cite{48} proposed transformer-based Rhythm2Drum to autoregressively generate drum onsets, which are then fed into a Unet to generate velocity and offset. After generating the drum track, the REMI representation \cite{95} and a transformer-XL-based \cite{96} encoder-decoder architecture are adopted to generate additional piano or guitar instruments. Similarly, Han et al. \cite{51} first utilized a transformer model to generate drum rhythm given dance movements and dance style, and then introduced the BERT-like \cite{61} model to enrich the remaining music track based on the drum track. The InteractiveBeat proposed by Liu et al. \cite{53} generated interactive rhythmic audio based on human movements, where a transformer-based DrumGenNet is proposed for synthesizing polyphonic drum tracks based on the estimated tempo and drum `style'. In contrast to the Rhythm2Drum, DrumGenNet discards the UNet that generates velocity and offset matrices and uses the continuous 1D rhythm input as velocity, yielding a more compact and real-time compatible drum generation network.

Based on compound words (CP)-like music representation \cite{97}, Di et al. \cite{38} and Tian et al.\cite{90} propose the controllable music transformer (CMT) models for generating music for arbitrary video. V-MusProd \cite{12} decouples music generation into three progressive transformer stages: chord, melody, and accompaniment, each conditioned on video features and the outputs of preceding stages. Kang et al. \cite{58} proposed an Affective Multimodal Transformer (AMT), feeding video embedding into the transformer encoder and chord embedding concatenated with the key into the transformer decoder to generate chord sequences autoregressively. \textcolor{black}{Sulun et al. \cite{163} developed a video-based MIDI generator, which adopted the event-based music representation and transformer to autoregressively generate music conditioned on emotions and scene boundaries extracted from the input video.} To ensure rhythmic alignment and stylistic matching, Liang et al. \cite{77} introduced a two-stage autoregressive encoder-decoder generation pipeline to employ the dance kinematic rhythm and dance style features, avoiding mutual interference. In the first stage, the Drum Transformer generates a drum track conditioned on predicted music beats. In the second stage, the Multi-track Transformer creates a multi-track MIDI conditioned on previously generated drum track and dance style features. 

However, symbolic music generation is often limited to a few classical instruments, and symbolic representation is not expressive enough to cover the diverse range of sounds we hear in typical soundtracks, which yields highly formulated results and hinders the model from synthesizing complex and diverse music. Moreover, the fidelity of the generated music is largely contingent upon the quality of a sound synthesizer or a MIDI playback engine, which may not adequately reflect the full depth and complexity of the actual instruments. Lastly, the small scale and limited genre diversity of MIDI annotations typically lead to poor generalization \cite{40}.

\paragraph{Audio}
When employing the autoregressive paradigm for audio generation, directly processing long-term music from the raw waveform is computationally intensive and challenging \cite{72}. Consequently, existing research typically first encodes the audio into discrete tokens using an audio compression algorithm, such as SoundStream \cite{98}, EnCodec \cite{99}, and Improved RVQGAN \cite{100}. Subsequently, an autoregressive sequence modeling approach is used to predict the discrete tokens, which are then reconstructed back into audio through the codec's audio decoder. In such approaches, the audio quality of the generated music is limited by the capabilities of the codec and substantial computational resources are required to train the music token prediction model.

To bridge the gap between the coarser video representation and the high-resolution audio representation, Su et al. \cite{78} proposed V2Meow, a high-fidelity music waveform generator conditioned on diverse video inputs. V2Meow follows the AudioLM \cite{33} pipeline, comprising three autoregressive stages: a transformer predicts music semantic tokens from visual features (and optional MuLan embeddings), two decoder-only transformers map semantic to coarse and fine acoustic tokens, and the SoundStream decoder reconstructs the audio. \textcolor{black}{The combination of Encodec and MusicGen has been widely adopted for V2M generation.} Tian et al. \cite{4}, Lin et al. \cite{40}, and Zuo et al. \cite{64} transform the audio into discrete tokens with EnCodec and use a transformer architecture to predict music tokens autoregressively. \textcolor{black}{Wu et al. \cite{159} also adopted Encodec as the default compression model and utilized a decoder-only transformer with devised temporal masks to restrict each music token to attend to only nearby video frames, enabling hierarchical alignment from global to local contexts and improving rhythm-aware correspondence. Rhythm-Controllable MusicGen \cite{155} integrates a rhythm conditioner, T5 text encoder, and MusicGen decoder to autoregressively generate music codebook projections. DyViM \cite{170} similarly adopts MusicGen as the decoder and EnCodec for waveform tokenization. You et al. \cite{171} first compress the predicted music features with a Conv1D layer. Then, a decoder-style cross-attention module generates music embeddings, where queries are the shifted real music embeddings, and keys/values are the predicted music features. The embeddings are optimized with RVQ codebook cross-entropy, and a frozen Music decoder converts them into audio.}

Although pre-trained text-to-music models \cite{34,101} demonstrate the ability to generate high-quality and diverse music, most current models only provide control over global attributes of the music and cannot manipulate dynamic properties such as rhythm. To address this issue, Li et al. \cite{15} employed a two-stage training model named VidMusician, which is built upon the MusicGen \cite{34}. The first stage introduces semantic features to capture global video content, while the second stage incorporates rhythmic cues using zero and identity initialization to ensure smooth integration without compromising the progress achieved in the first stage. Li et al. \cite{17} learned two pseudo-words that represent rhythm and genres in dance videos, and then integrated them into the vocabulary of the text encoder. The encoded text embedding can condition various text-to-music backbones, such as transformer-based MusicGen \cite{34} as well as diffusion-based Riffusion \cite{101} and AudioLDM \cite{102}. \textcolor{black}{Guan et al. \cite{165} first generated the textual description of music based on the visual condition and then used the pretrained MusicGen to generate the corresponding music. Without paired data, the mapping between visual description and musical description is achieved through three steps of Chain-of-Thought (CoT) reasoning by the LLM. Specifically, they proposed a two-stage approach, where the first stage employs an MLP to map CLIP-based visual data to the LLM2Vec-based \cite{166} reasoning process (aka. latent information), and the second stage uses an embedding diffusion model (EmbedDiff) to map latent information to the T5-based music embeddings.} Recent MLLMs \cite{65,70} similarly adopt the pretrained MusicGen for cross-modal music generation. However, they use the MUVideo dataset curated by gathering music samples from the AudioSet \cite{103}. Since the samples in AudioSet are all 10 seconds, these models perform particularly well on 10-second videos.

\subsubsection{Non-autoregressive Modeling.}
\paragraph{Symbolic}
Given five consecutive grayscale video frames, Koepke et al. \cite{13} and Su et al. \cite{14} predicted all note onsets occurring around the middle video frame. Koepke et al. \cite{13} passed all outputs of a given video through a Gaussian filter to add temporal smoothing and thresholded them to obtain per-note binary outputs for MIDI conversion. Su et al. \cite{14} trained a GAN \cite{104} to refine the predicted results toward the Pseudo GT MIDI obtained using the Onset and Frames framework \cite{105}, and then a deep-learning-based MIDI synthesizer converts MIDI to the spectrogram. Li et al. \cite{59} first converted the original MIDI files into piano roll representations and employed a latent diffusion model as the generator to produce piano rolls conditioned on videos. The generated piano rolls are then converted into music using a music processor.
\paragraph{Audio}
Classic generative models such as VAE and GAN are initially used for V2M generation. To generate multi-instrumental music from videos in an unsupervised setting, Su et al. \cite{45} introduced a multi-band residual 1D convolutional VQ-VAE for learning a discrete latent representation of various instrumental music. Post training, the autoregressive prior network first generates discrete latent representations conditioned on the musicians' movement features. Then, the discrete latent representations were decoded by the VQ-VAE decoder to synthesize instrument-specific music. To tackle the challenge of high dimensionality of audio data, Zhu et al. \cite{54} proposed D2M-GAN, an adversarial multi-modal framework that generates complex musical samples conditioned on dance videos. The convolution-based VQ generator takes the fused motion-visual data as input and outputs the desired VQ audio representations. The multi-scale discriminator evaluates the generated and real VQ representations. The generated VQ codes were mapped through a pre-learned codebook and decoded into raw audio using the JukeBox decoder \cite{106}. However, the high variability of waveform data (e.g., variable and high-dynamic phase, energy, and timbre of instruments) makes it difficult to directly model high-quality waveforms. As a result, the generated music often contains strange noises \cite{51}. Moreover, the slow decoding speed of the VQ audio representation results in short-length music (2~6s) \cite{11,72}. 

The emergence of diffusion probabilistic models (DPMs) \cite{92,107} has revolutionized the content generation paradigm. Inspired by recent advances in the computer vision, studies have adapted diffusion models as music generators. Rather than operating directly on high-dimensional raw data, latent diffusion models (LDMs) work in a compressed latent space that preserves core structure and semantic information. The process begins with an encoder, such as a Variational Autoencoder (VAE) or Vector Quantized VAE (VQ-VAE), which maps the raw data into a lower-dimensional representation. Diffusion processes are then applied in the latent space, making the generative modeling more computationally efficient due to the reduced dimensionality. Once the generation process is complete, a decoder reconstructs the latent representation back into its original form.

\textcolor{black}{The combination of VAE with latent-space diffusion models (LDM) or flow-matching models is widely used for V2M generation.} As the follow-up work of D2M-GAN \cite{54}, Zhu et al. \cite{55} proposed a contrastive diffusion framework that combines latent diffusion training and contrastive learning \cite{108} to achieve diverse multimodal conditional synthesis tasks including dance-to-music generation. Instead of relying on continuous latent spaces, they train a VQ-based model to obtain discrete music representations. Then, they introduce a Conditional Discrete Contrastive Diffusion (CDCD) loss and design two contrastive diffusion mechanisms to effectively incorporate this loss into the denoising process. One is the step-wise parallel diffusion with intra-negative samples, which captures intra-sample correlations such as rhythm, and the other is the sample-wise auxiliary diffusion with inter-negative samples, which enhances instance-level attributes such as genre. Tan et al. \cite{75} leveraged a combination of a UNET-based latent diffusion model and a pre-trained VAE model to achieve 3D motion to dance music generation. However, their model performs well only for generating 5-second audio samples. Extending the output to longer sequences results in a decrease in audio quality. In contrast, LORIS \cite{72} proposed by Yu et al. can generate long-term waveforms (25s~50s) with affordable costs. LORIS adopts a latent diffusion probabilistic model with a hierarchical cross-attention module to serially attend visual, rhythm, and genre conditionings. Zhang et al. \cite{11} first used the pretrained Diffusion Magnitude Autoencoder (DMAE) \cite{109} to compress music into latent space, and then proposed a U-Net-based dance-to-music diffusion generator to generate the latent music representation. Li et al. \cite{82} adopted the flow-matching based music generation and utilized a pretrained unconditional DiT generator to inherit its generative capabilities. The classifier-free guidance (CFG) \cite{117} is utilized for unconditional or conditional generation. Zhang et al. \cite{16} combined a VAE and a diffusion U-Net model derived from Stable Audio's framework \cite{110} for video background music generation. \textcolor{black}{Other extensions include a dual-diffusion mechanism that jointly exploits positive and negative conditioning \cite{152}, a conditional soundtrack generator combining AudioMAE with UNet diffusion for joint music and sound-effect synthesis \cite{157}, and MotionComposer \cite{158}, a two-stage diffusion framework that first predicts beats (BeatNet) and then composes chords (ChordNet) using the visual embedding, retrieved chord reference, and the predicted music beat from BeatNet. Additionally, the recent VeM \cite{172}, Diff-V2M \cite{173}, and GACA-DiT \cite{174} all take the LDMs or latent flow matching as the generative backbone for V2M generation. }

Not specifically for the V2M generation, Ruan et al. \cite{31} proposed the first multi-modal diffusion model (MM-Diffusion) for joint audio-video generation. It adopts a coupled U-Net architecture with separate U-Nets for audio and video, and employs a multimodal attention mechanism with random-shift attention masks to efficiently align the two modalities. Although MM-Diffusion is designed for joint audio-video generation, it can achieve video-to-audio in a zero-shot transfer manner via replacement-based or gradient-guided methods. \textcolor{black}{Similarly, MM-LDM \cite{168} is an audio-video joint generation model based on two LDMs specifically for video and audio, respectively. It also exhibits remarkable adaptability when extended to video-to-audio generation.} The MoMu-Diffusion proposed by You et al. \cite{111} also enables multiple tasks including motion-to-music, music-to-motion, and joint motion-music generation, which adopts the transformer-based diffusion model. 

\textcolor{black}{Furthermore, approaches encompassing multimodality-to-music, multimodality-to-multiaudio, and multimodality-to-multimodality paradigms can also achieve V2M generation. Visual Music Bridge (VMB) \cite{160} can produce music from diverse input modalities, including text, videos, and images, which integrates two explicit bridges (i.e., text bridge and retrieved music bridge) into a text-to-music diffusion transformer, i.e., the stable audio open \cite{178}. Recent MLLMs such as MuMu-LLaMA \cite{188}  and Next-GPT \cite{113}} adopted diffusion-based models including AudioLDM \cite{102}, and AudioLDM 2 \cite{114} for cross-modal audio generation. Other multimodal diffusion models like CoDi \cite{112} and diffusion latent aligner \cite{115}, although not specifically designed for multimodal music understanding or generation tasks, are capable of V2M generation. However, their limited music training data restricts their musical capabilities. \textcolor{black}{AudioX \cite{167} is a unified Diffusion Transformer model for anything-to-audio and music generation, which supports a broad range of input conditions, including text, image, video, and audio within a single framework. It integrates specialized encoders for video, text, and audio and adopts a multi-modal masked training strategy to encourage robust cross-modal interactions and enhance representation learning. AudioGenie \cite{169} is a novel training-free multi-agent system, which covers a comprehensive range of input modalities (text, image, video, and combination) and audio types (sound effects, speech, song, music, and combination) in a unified framework.}

\textcolor{black}{All2All multimodal foundation models aim to establish a unified framework for multimodal understanding and generation, enabling flexible, bidirectional mappings and joint generation across diverse modalities such as text, images, audio, and video. In this sense, the relationship between All2All models and V2M generation can be viewed as that between a general paradigm and a specific application. Unlike All2All models, which prioritize universality and holistic capability, V2M tasks emphasize targeted modeling and domain-specific adaptation, with higher requirements for rhythm synchronization, attribute controllability, etc. While All2All models provide powerful general representations and strong transfer potential, making them suitable as backbones or pretraining foundations for V2M, they typically underperform in fine-grained controllability and thus require task-specific modules and domain-adaptive fine-tuning to meet the specialized demands of V2M generation.}

\subsubsection{\textcolor{black}{Summary}}
\textcolor{black}{
Based on the above survey, in the field of V2M generation, audio has increasingly surpassed symbolic music as the dominant output format. This trend is largely attributed to the development of large-scale models and the relative ease of collecting large audio datasets. Moreover, audio is generally more expressive, conveying richer musical nuances without requiring additional post-processing steps such as MIDI synthesis or rendering. Both autoregressive and non-autoregressive approaches remain widely used, despite inherent trade-offs between them. Autoregressive models exhibit slower inference but are capable of preserving long-range musical structure. Non-autoregressive models, on the other hand, tend to produce audio with higher fidelity and richer timbral detail. In particular, flow-matching methods offer more stable training and require fewer sampling steps compared to diffusion-based approaches, resulting in substantially faster inference relative to autoregressive models.}

\begin{table}[]
\caption{The comparison for video-to-music generation methods. Note that this table includes \textbf{ONLY} methods with available demos, to enable a fair comparison of generation quality.}
\label{tab2}
\begin{tabular}{|l|l|p{1.2cm}|p{1.2cm}|l|}
\hline
Method & Demo & Music Length & Model Size & Training Resources  \\ \hline
\multicolumn{5}{|c|}{\cellcolor{lightgray!25}Slient Performance}     \\ \hline
Audeo \cite{14} & \href{https://faculty.washington.edu/shlizee/audeo/}{Link}     &  30s &-   &2 Titan X GPUs   \\ \hline
Foley Music \cite{10} &\href{http://foley-music.csail.mit.edu/}{Link}  &  6s   &- &- \\ \hline
MI Net \cite{45} & \href{https://www.youtube.com/watch?v=yo5OZKBbBh4}{Link}  &  4s  &- &- \\ \hline
\multicolumn{5}{|c|}{\cellcolor{lightgray!25}Dance or Sports}     \\ \hline
Dance2Music \cite{46} & \href{https://sites.google.com/view/dance2music/home}{Link}     &  \textasciitilde12s   &3M   & -    \\ \hline
Rhythmic-Net \cite{48} & \href{https://github.com/shlizee/RhythmicNet}{Link}     &  16 bars   &359M   & 2 Titan X GPUs   \\ \hline
D2M-GAN \cite{54} & \href{https://l-yezhu.github.io/D2M-GAN/}{Link}     &  2s   &535M   & -   \\ \hline
CDCD \cite{55} & \href{https://l-yezhu.github.io/CDCD/}{Link}     &  6s   &6.4G   &4 NVIDIA
RTX A5000 GPUs     \\ \hline
LORIS \cite{72} & \href{https://justinyuu.github.io/LORIS/}{Link}     &  25s or 50s   &3.12G   &8 NVIDIA A100 GPUs  \\ \hline
Motion to Dance \cite{75}  & \href{https://dmdproject.github.io/}{Link}     &  5s &846M   &1 NVIDIA RTX A5000 GPU    \\ \hline
Dance2MIDI \cite{51} & \href{https://dance2midi.github.io/}{Link}     &30s &-   &-    \\ \hline
Textual Inversion \cite{17} & \href{https://www.youtube.com/watch?v=y2pG2S5xDLY&feature=youtu.be}{Link}     &5.12s &\textasciitilde25G   &1 NVIDIA A40 GPU    \\ \hline
Dance2Music-Diffusion \cite{11} & \href{https://www.youtube.com/watch?v=eCvLdLdkX-Y}{Link}     &5.9s &\textasciitilde900M   &1 NVIDIA RTX 2080ti GPU 
(11GB)   \\ \hline
MoMo-Diffusion \cite{111} & \href{https://momu-diffusion.github.io/}{Link}     &25s or 50s &-   &8 NVIDIA 4090 GPUs  \\ \hline
MotionComposer \cite{158} & \href{https://beria-moon.github.io/Soundtrack-your-Motion/}{Link}     &25s &-   &1 NVIDIA A6000 GPU  \\ \hline
GACA-DiT \cite{174} & \href{https://beria-moon.github.io/GACA-DiT/}{Link}     &5s &-   &-    \\ \hline
\multicolumn{5}{|c|}{\cellcolor{lightgray!25}General}     \\ \hline
CMT \cite{38} & \href{https://wzk1015.github.io/cmt/}{Link}   &3min  &188M  &4 RTX 1080Ti GPUs   \\ \hline
V-MusProd \cite{12} & \href{https://drive.google.com/drive/folders/1ASY44xqWGZgKkcHhpzWlOhIbUIMe_epQ}{Link}   &6min  &-  &1 V100 GPU   \\ \hline
MuMu-LLaMA (M\textsuperscript{2}UGen) \cite{65} & \href{https://crypto-code.github.io/MuMu-LLaMA_Demo/}{Link}   &30s  &\textasciitilde25G  &- \\ \hline
Video2Music \cite{58} & \href{https://amaai-lab.github.io/Video2Music/}{Link} &300s  &125M  &NVIDIA Tesla V100 DGXS 32 GB GPUs    \\ \hline
Diff-BGM \cite{59} & \href{https://github.com/sizhelee/Diff-BGM/blob/master/video.mp4}{Link} &8 bars  &506M  &-   \\ \hline
V2Meow \cite{78} & \href{https://google-research.github.io/noise2music/v2meow/}{Link} &10s  &-  &-    \\ \hline
VidMuse \cite{4} & \href{https://vidmuse.github.io/}{Link} &30s  &7.87G  &64 H800 GPUs  \\ \hline
MuVi \cite{82} & \href{https://muvi-v2m.github.io/}{Link} &4\textasciitilde30s  &-  &8 NVIDIA V100 GPUs  \\ \hline
VMAS \cite{40} & \href{https://genjib.github.io/project_page/VMAs/index.html}{Link} &10s  &-  &32 NVIDIA GPUs (4000G) \\ \hline
VidMusician \cite{15} & \href{youtu.be/EPOSXwtl1jw}{Link} &30s  &\textasciitilde100M  &4 NVIDIA Tesla V100-SXM2 GPUs \\ \hline
SONIQUE \cite{16} & \href{https://zxxwxyyy.github.io/templates/sonique.html}{Link} &23.8s  &33G  &4 RTX8000 GPUs \\ \hline
Mozart's touch \cite{70} & \href{https://tiffanyblews.github.io/MozartsTouch-demo/}{Link} &10s  &\textasciitilde25G  &1 NVIDIA RTX 3090 24GB GPU \\ \hline
Visuals Music Bridge (VMB) \cite{160} & \href{https://wzk1015.github.io/vmb/}{Link} &30s  &-  &4 NVIDIA A100 GPUs  \\ \hline
XMusic \cite{90} & \href{https://sites.google.com/view/xmusicdemos}{Link} &30s-2min  &-  &8 NVIDIA A800 GPUs \\ \hline
EMSYNC \cite{163} & \href{https://em-sync.github.io/}{Link} &1min  &\textasciitilde150M  &1 NVIDIA A100 80GB GPU \\ \hline
CoT-VTM \cite{165} & \href{https://xxkkxxx.github.io/cot-vtm/}{Link} &10s  &-  &1 NVIDIA RTX A6000 GPU  \\ \hline
FilmComposer \cite{155} & \href{https://apple-jun.github.io/FilmComposer.github.io/}{Link} &30s  &-  &1 NVIDIA A6000 GPU \\ \hline
VeM \cite{172} & \href{https://vem-paper.github.io/VeM-page/}{Link} &10-60s  &-  &1 NVIDIA A100 GPU  \\ \hline
Diff-V2M \cite{173} & \href{https://tayjsl97.github.io/Diff-V2M-Demo/}{Link} &30s  &23.4G  &2 NVIDIA A100 GPUs  \\ \hline
\end{tabular}
\end{table}
\subsection{\textcolor{black}{Method Comparisons}}
\textcolor{black}{
Direct quantitative comparison across existing V2M generation systems is challenging, as most methods are trained and evaluated on different datasets, use distinct preprocessing pipelines, or report inconsistent metric values across papers. These discrepancies make it difficult to consolidate results into a unified benchmark or identify a definitive state-of-the-art method. To address this limitation, this section presents public demo links (when available) of existing methods to facilitate subjective assessment by readers. Beyond generation quality, we compare methods from multiple practical and engineering-oriented dimensions, e.g., inference latency, training requirements, as well as model size and computational footprint. Table \ref{tab2} lists demos, generation lengths, model sizes, and training resources to support transparent and intuitive comparisons. Since exhaustive one-by-one comparisons are impractical, the subsequent discussion provides a high-level analysis contrasting autoregressive and non-autoregressive approaches.}

\subsubsection{\textcolor{black}{Training}}
\textcolor{black}{
Different video-to-music generation paradigms exhibit distinct levels of data and computational demand. Symbolic music models generally require smaller architectures and noticeably lighter training resources, benefiting from the compactness of symbolic representations compared to audio with high sampling rates. Autoregressive audio token models, such as VidMuse, VMAs, often involve significant memory consumption and long attention spans, which impose substantial computational costs. In contrast, non-autoregressive approaches, particularly diffusion and flow-matching based models, introduce additional latent-space learning and multi-stage training. Diffusion frameworks tend to have higher training complexity than AR methods, while flow-matching variants benefit from more stable optimization and faster convergence, offering a more computationally efficient alternative without sacrificing expressiveness.}

\subsubsection{\textcolor{black}{Inference}}
\textcolor{black}{
Inference behavior also varies considerably across architectures. Autoregressive methods generate music token by token, leading to latency that scales with output length. Although techniques such as KV-cache, chunked attention, and hierarchical tokenization can partially mitigate this, autoregressive models remain inherently sequential. Diffusion models require iterative denoising across many sampling steps, resulting in relatively high latency for each audio window despite their strong fidelity and detail. This makes real-time or interactive usage challenging without efficient sampling acceleration. In contrast, flow-matching models offer faster inference due to reduced sampling steps and more efficient parameterization of the generation trajectory. Consequently, flow-matching models provide an improved trade-off between latency and audio quality, especially for interactive or low-latency video-music applications.}

\subsubsection{\textcolor{black}{Model Size}}
\textcolor{black}{
Model size and computational footprint further differentiate these approaches. Symbolic music generation systems remain the most lightweight, enabling efficient training and scaling to long musical durations with minimal hardware requirements. Autoregressive audio models typically occupy a moderate range in terms of parameters and storage, but their sequential decoding introduces a substantial computational burden during inference, particularly for high-resolution audio. Diffusion and flow-matching architectures generally involve larger model backbones, resulting in higher memory and storage costs. Nevertheless, these larger models often achieve superior timbral richness and acoustic details. Additionally, models based on text-to-music foundations, such as MusicGen, AudioLDM, and Stable Audio, generally exhibit considerably large model sizes due to the high-capacity backbone architectures.}
\begin{table}[]
\caption{Dataset Summary. Note that this table lists \textbf{ONLY} datasets with publicly accessible downloads.}
\label{tab3}
\begin{tabular}{|p{1.5cm}|l|p{1.5cm}|p{1.5cm}|p{3.8cm}|l|l|}
\hline
Dataset & Year & Video & Music & Annotation &Size & Access  \\ \hline
URMP \cite{119}	&2018 &Silent performance &Symbolic \& Audio &Instrument, Note annotation, Pitch contour &44 (1.3h)	&\href{https://datadryad.org/stash/dataset/doi:10.5061/dryad.ng3r749}{Link} \\ 
MUSIC \cite{120}	&2018 &Silent performance &Symbolic &Instrument &714 (~23.8h)	&\href{https://github.com/roudimit/MUSIC_dataset}{Link} \\ 
CoP Benchmark \cite{162}	&2025 &Silent performance &Symbolic \& Audio &- &10 hours &\href{https://github.com/acappemin/Video-to-Audio-and-Piano}{Link}   \\ \hline
AIST \cite{121}	&2019 &Dance &Audio &Dance genre, Dancer, Camera, Choreography &13,940 (118.1h)	&\href{https://aistdancedb.ongaaccel.jp/database_download}{Link} \\ 
AIST++ \cite{122} &2021 &Dance &Audio &Dance genre, Dancer, Camera, Choreography &1,408 (5.2h)	&\href{https://google.github.io/aistplusplus_dataset/download.html}{Link} \\ 
TikTok Dance-Music \cite{54}	 &2022	&Dance &Audio &-	&445 (1.5h)	&\href{https://github.com/L-YeZhu/D2M-GAN?tab=readme-ov-file}{Link}  \\ 
LORIS \cite{72}	&2023 &Dance, Floor exercise, Figure skating	&Audio &Genre &16,593 (86.43h)	&\href{https://huggingface.co/datasets/OpenGVLab/LORIS}{Link}  \\ 
InDV \cite{17} &2024	&Dance	&Audio &Dance genre	&595 (0.85h)	&\href{https://drive.google.com/file/d/1UlhzJ6a7Xox5ebTu74ZNb5jCK6yK2RwQ/view}{Link}  \\ \hline
HIMV-200K \cite{130} &2018 &General &Audio &Visual entities &200,500 &\href{https://github.com/csehong/VM-NET/tree/master/data}{Link} \\ 
MusicCaps \cite{126} &2023 &General &Audio &Musical caption, Musical aspects: genre, mood, tempo, singer voices, instrumentation, dissonances, rhythm, etc.  &5,521 (15.34h)	&\href{https://www.kaggle.com/datasets/googleai/musiccaps}{Link} \\ 
SymMV \cite{12} &2023 &General: MV &Symbolic \& Audio &Genre, Lyrics	&1,181 (78.9h)	&\href{https://github.com/zhuole1025/SymMV/tree/main/dataset}{Link} \\ 
MuVi-Sync \cite{58} &2024 &General: MV &Symbolic \& Audio &Scene offset, Emotion,
Motion, Semantic, Chord, Key, Loudness, Note density &784	&\href{https://zenodo.org/records/10057093}{Link} \\ 
Popular Hooks \cite{127} &2024 &General: MV &Symbolic \& Audio &Lyrics, Tonality, Chord progression, Structure, Genre, Emotion, Region &38,694 (249h)	&\href{https://huggingface.co/datasets/NEXTLab-ZJU/popular-hook}{Link} \\ 
BGM909 \cite{59} &2024 &General: MV &Symbolic \& Audio &Chords, Beats, Key signatures, Language descriptions, Shot transitions, Lyrics, Genre, Rhythmic patterns &909	&\href{https://github.com/sizhelee/Diff-BGM}{Link} \\ 
V2M \cite{4} &2024 &General &Audio &Video genre &360k (18,000h) &\href{https://huggingface.co/datasets/HKUSTAudio/VidMuse-V2M-Dataset}{Link} \\ 
MusicPro-7k \cite{155} &2025 &General: Film &Audio &Description, Rhythm spots, Main melody 
&7,418	&\href{https://huggingface.co/datasets/apple-jun/MusicPro-7k}{Link} \\  
HarmonySet \cite{164} &2025 &General &Audio &Video genre, Key moments, Content alignment, Narrative enhancement, Detailed descriptions of the video-music relationship  &48,328
(458.8h)	&\href{https://huggingface.co/datasets/Zzitang/HarmonySet}{Link} \\ 
OSSL \cite{175} &2025 &General: Film &Audio &Mood, Movie genre &736 (36.5h)	&\href{https://havenpersona.github.io/ossl-v1/}{Link} \\ \hline 
\end{tabular}
\end{table}
\section{DATASETS}
This section categorizes existing multimodal video-music datasets from the perspective of video types. The brief introductions and access to these datasets are summarized in Table \ref{tab3}.

\subsection{Human-centric}
\subsubsection{Silent Performance.}
\textbf{The University of Rochester Multimodal Music Performance (URMP) dataset} \cite{119} comprises 44 simple multi-instrument classical music pieces recorded in a studio. For each piece, the dataset includes video recordings of performances, MIDI files, individual and mixed audio recordings in WAV format, as well as frame-level pitch trajectories and note-level transcriptions for individual tracks in ASCII-delimited text format.

\textbf{MUSIC (Multimodal Sources of Instrument Combinations) dataset} \cite{120} is an untrimmed video dataset downloaded by querying keywords from YouTube. It contains 714 music performance videos belonging to 11 instrument categories, including 565 solo videos and 149 duet videos. The average duration of each video is approximately 2 minutes.

\textcolor{black}{\textbf{CoP Benchmark Dataset} \cite{162} is a fully open-sourced, multi-modal benchmark designed specifically for video-guided piano music generation. It features 1) detailed multi-modal annotations: comprehensive labels facilitate precise semantic and temporal alignment between video content and piano audio, leveraging step-by-step (Chain-of-Perform) guidance. 2) Versatile evaluation framework: it enables rigorous evaluation of both general-purpose and specialized piano music generation tasks from videos. The primary constraint for data collection was a five-view piano performance with a fully visible keyboard and practice pedal.}

\subsubsection{Dance.}
\textbf{AIST} dance video database\footnote{https://aistdancedb.ongaaccel.jp/} \cite{121} consists of 1,618 street dances among 13,940 videos, including 13,888 videos of 10 street dance genres and 52 additional videos with three different situations. This database includes 10 street dance genres, 35 dancers (20 male, 15 female), 9 camera viewpoints, and 60 musical pieces covering 12 types of tempo, a total of 118.1 hours.

\textbf{AIST++} Dance Motion Dataset\footnote{https://google.github.io/aistplusplus\_dataset/factsfigures.html} \cite{122} is constructed from the AIST Dance Video database, which is a 3D human dance motion-music paired dataset, containing 1408 sequences of 3D human dance motion represented with the SMPL format \cite{42}. The dance motions are equally distributed among 10 dance genres with hundreds of choreographies. Motion durations vary from 7.4 sec. to 48.0 sec, a total of 5.2 hours. All the dance motions have corresponding music. 

\textbf{TikTok Dance-Music dataset} \cite{54} consists of 445 "in-the-wild" dance videos, each with an average length of 12.5 seconds. This dataset features 85 different songs, with the majority of videos showcasing a single dance performer, and a maximum of five performers. Unlike AIST, this dataset presents a greater challenge and more accurately reflects real-world scenarios, establishing a new benchmark for future research. Zhu et al. \cite{54} provided the full dataset annotation including the video link, music ID, video ID, the duration, and the number of dancers here, as well as their splits for training and testing. To facilitate the usage, preprocessed motion, vision, and music audio are also available.

\textbf{D2MIDI} \cite{51} is the first multi-instrument MIDI and dance paired dataset, which encompasses six mainstream dance genres: classical, Hip-Hop, ballet, modern, Latin, and house, and 13 MIDI instrument types: Acoustic Grand Piano, Celesta, Drawbar Organ, Acoustic Guitar (nylon), Acoustic Bass, Violin, String Ensemble 1, SynthBrass 1, Soprano Sax, Piccolo, Lead 1 (square), Pad 1 (new age), and Drum. This dataset is large-scale in that it contains 71,754 30s dance videos in total. For dance motion, 3D keypoints of the human body including body, hand, and face are extracted, allowing densely represented pose features.

\textbf{LORIS} dataset \cite{72} is a large-scale rhythmic video soundtrack dataset that includes 86.43h long-term, high-quality raw videos with corresponding 2D poses, RGB features, and ameliorated audio waveforms. This dataset is originally used for the video background music generation task. There are 12,446 25-second paired videos, including 1,881 dancing videos, 8,585 figure skating videos, and 1,950 floor exercise videos. For the 50-second versions, the LORIS dataset includes 4,147 figure skating videos and 660 floor exercise videos. The dancing videos are curated from AIST++, figure skating videos are collected from FisV \cite{123} dataset and FS1000 \cite{124} dataset, and floor exercise videos are from Finegym \cite{125} dataset.

\textbf{InDV (In-the-wild Dance Videos)} dataset \cite{17} contains 595 5.12-second clips with 216 songs. The dataset is categorized into 10 genres: Ballet, Breaking, Chinese Traditional Dance, Hip-hop, Jazz, Latin, Locking, Popping, and Waacking.

\subsection{General}
\textbf{Hong \& Im Music–Video 200K (HIMV-200K)} \cite{130} is a benchmark dataset used for video–music retrieval, composed of 200,500 data pairs. HIMV-200K includes official music videos, parody music videos, and user-generated videos. These video–music pairs are from the YouTube-8M\footnote{https://research.google.com/youtube8m/} \cite{131}, a large-scale video classification benchmark that consists of millions of YouTube video IDs and associated labels. Despite its size, it suffers from weak correspondence and poor quality, e.g., static videos, live performances, or amateur montages, since it is only designed for the retrieval task \cite{12}.

\textbf{MusicCaps} dataset \cite{126} contains 5,521 music examples, each of which is labeled with an English aspect list, a free text caption written by musicians, the YouTube video ID, and the times the labeled music segment appears. All the examples are 10-second music clips from the AudioSet dataset\footnote{ https://research.google.com/audioset/} \cite{103}.

\textbf{Symbolic Music Videos (SymMV)} \cite{12} is a video and symbolic music dataset curated from YouTube, along with chord, rhythm, tonality, melody, accompaniment, and other metadata such as genre and lyrics. Overall, it contains 1181 video-music pairs of more than 10 genres with a total length of 78.9 hours.

\textbf{MuVi-Sync} \cite{58} is an openly available music video dataset that consists of video features (scene offset, emotion, motion, and semantic) and music features (chord, key, loudness, and note density) extracted from a total of 748 music videos curated from YouTube. For the calculation of different features, please refer to \cite{58}. 

\textbf{Popular Hooks} \cite{127} is a publicly accessible multimodal music dataset comprising 38,694 popular musical hooks with synchronized MIDI, music video, audio, and lyrics. Furthermore, the dataset provides detailed annotations of high-level musical attributes such as tonality, chord progression, structure, genre, emotion, and region. The multimodal music data within the Popular Hooks dataset is accurately time-aligned. Emotion categories follow Russell's 4Q \cite{128}.

\textbf{BGM909} \cite{59} is a high-quality video-music dataset that includes extensive music files with detailed annotations such as chords, beats, key signatures, and official music videos. The video-music pairs have been manually edited and checked to ensure perfect temporal alignment. Each video is accompanied by fine-grained natural language descriptions, video shot transitions, and additional metadata, including lyrics, genre, and rhythmic patterns. The style for each sample is predicted by GPT. BGM909 is based on POP909 \cite{129}, which consists of 909 piano versions of music along with corresponding well-aligned videos.

\textcolor{black}{
\textbf{V2M} dataset \cite{4} comprises 360K pairs of videos and music, collected from YouTube and covering various types including movie trailers, advertisements, and documentaries. It contains three subsets: V2M-360K for pretraining (\textasciitilde18000h), V2M-20K for finetuning (\textasciitilde600h), and V2M-bench (300 pairs, \textasciitilde9h) for evaluation.}

\textcolor{black}{
\textbf{MusicPro-7k} \cite{155} is a professional, versatile, and high-quality video-to-music Dataset, specifically focused on film music, featuring about 7,418 samples, each with the film clip, high-quality music, visual description, music description, main melody, and rhythm spots. 7K Film clips are derived from LSMDC \cite{156}, selecting video segments over 8 seconds long from numerous classic films. The source films cover all common genres, including drama, thriller, comedy, romance, and more.}

\textcolor{black}{
\textbf{HarmonySet} \cite{164} is the first instruction tuning dataset for MLLMs to understand the alignment between video and music, which consists of 48,328 diverse video-music pairs, annotated with detailed information on rhythmic synchronization, emotional alignment, thematic coherence, and cultural relevance. Even though this dataset was collected to facilitate video–music understanding, it can also be employed for V2M generation tasks.}

\textcolor{black}{
\textbf{Open Screen Soundtrack Library (OSSL)}\cite{175} is a dataset consisting of movie clips from public domain films, totaling approximately 36.5 hours, paired with high-quality soundtracks and human-annotated mood information. The frame rate and resolution for each video are 25fps and 960$\times$720, respectively, and the sampling rate for each soundtrack is 44.1kHz. Mood is classified into four categories based on Russell's 4Q \cite{128}, i.e., HVHA (high valence, high arousal), HVLA (high valence, low arousal, LVHA (low valence, high arousal), and LVLA (low valence, low arousal).}

\textcolor{black}{
\textbf{Open Screen Soundtrack Library Version 2 (OSSL-v2)} \cite{176} is a dataset consisting of 552.7 hours and 76,408 video clips sourced from both public domain movies as well as commercial ones from a publicly available dataset \cite{177}. The source-separated soundtracks average 26.04 seconds in length, along with rich metadata such as movie genres, release year, and title.}

Additionally, many studies \cite{15,40,82} have opted to collect videos from public media platforms like YouTube as training datasets. However, these datasets have not yet been made publicly available.
\section{EVALUATION}
In the realm of artificial intelligence generative content (AIGC), evaluation typically encompasses both objective metrics and human subjective assessment to ensure a comprehensive understanding of model performance. The field of video-to-music generation is no exception to this paradigm. This section summarizes the common evaluation metrics in V2M generation.

\subsection{Objective}
The objective evaluation of V2M generation typically assesses the following aspects: 1) audio fidelity; 2) the diversity of the generated audio; 3) the semantic alignment between the given video and the generated music; and 4) the temporal synchronization between the given video and the generated music.

\paragraph{Fidelity.}
For symbolic music, some statistical descriptors \cite{38,48,51,59,77,90,132} are frequently used for evaluating the quality of generated music, such as Pitch Count per Bar (PCB), average Pitch Interval (PI), average Inter-Onset Interval (IOI), Pitch Class Histogram Entropy (PHE), Grooving Pattern Similarity (GS), Scale Consistency(SC), Empty Beat Rate (EBR). The objective metrics of symbolic music generation have been well summarized by Ji et al. \cite{27}. While for musical audio, several audio evaluation metrics have been widely adopted, as follows.

\textbf{Frechet Audio Distance (FAD)} \cite{75,78,126} evaluates audio quality by measuring the similarity between the generated audio and the ground truth. Lower FAD values indicate more plausible audio. The calculation of FAD relies on an audio embedding model, with some commonly utilized models including VGGish \cite{133} and PANNs \cite{134}.

\textbf{Kullback Leibler Divergence (KLD)} \cite{75,78} can be used to reflect the acoustic similarity between the generated and reference samples, computing over PANNs' multi-label class predictions \cite{4}. It can also measure the distance between the predicted class probabilities of the generated music and the ground truth genre labels, with the help of a pre-trained genre classifier, such as MS-SincResNet \cite{139} and LEAF classifier \cite{140}.

\textbf{Frechet Inception Distance (FID)} \cite{135} originally evaluates image quality in GANs and it measures the distance between InceptionV3 \cite{136} pre-classification feature distributions for real and generated samples. Liu et al. \cite{53} adapted InceptionV3 input for a 2D magnitude spectrogram and train it on the Groove MIDI dataset \cite{137} for classifying 12 drum genres. Then, FID is computed from the extracted 2048-sized vectors from the last layer for both sets of samples.

\textbf{Density} \cite{4,138} measures how closely the generated samples match the real ones, by rewarding samples in regions where real samples are densely packed, relaxing the vulnerability to outliers.

\textcolor{black}{\textbf{MuLan Cycle Consistency (MCC)} \cite{126} is used to determine whether generated music is semantically relevant to the reference audio or text. For the video to music task, Su et al. \cite{78} computed MCC as the average cosine similarity between the MuLan \cite{81} audio embeddings of the generated music audio and the ground truth audio. Similarly, \textbf{CLAP Score} \cite{15,17,63} is computed as the average cosine similarity between the CLAP embeddings of real and generated music.}

\paragraph{Diversity.}
\textbf{Number of Statistically-Different Bins (NDB)} \cite{10,48,53,132} is used to evaluate the diversity of generated sound. To compute NDB, the training examples are clustered into k Voronoi cells by the k-means algorithm. Each generated example in the testing set is assigned to the nearest cell. NDB indicated the number of cells in which the training samples are significantly different from the number of testing examples. For NDB, lower is better.

\textbf{Diversity.} Li et al. \cite{59} proposed a metric to evaluate the diversity of the generated music. They randomly divide the generated music into two subsets, and each set has the same number of samples. The diversity of the generated music is defined as the average Euclidean distance between the music features of corresponding samples in two subsets.

\textbf{Coverage} \cite{138} measures the fraction of real samples whose neighborhoods contain at least one fake sample, i.e., assesses whether the generated samples capture the full range of variation found in real samples \cite{4}. Coverage is bounded between 0 and 1.

\textcolor{black}{\textbf{Chroma-based Diversity} \cite{155} is calculated by averaging the pairwise differences between all distinct pairs:
$D = 1 - 
\frac{2}{N(N-1)}
\sum_{i=1}^{N}
\sum_{j=i+1}^{N}
{Similarity}(C_i, C_j),
$
where $N$ is the total number of chroma feature matrices,  
$C_i$ is the $i$-th chroma feature matrix in the set, and  
${Similarity}(C_i, C_j)$ is the overall similarity over all segments between $C_i$ and $C_j$.
}

\paragraph{Semantic Alignment.}
\textbf{Genre Accuracy Score} \cite{54,55} evaluates whether the generated music samples have a consistent genre with the dance style. The musical samples with the highest similarity scores are retrieved from the segment-level database formed by original audio samples with the same sequence length. The similarity scores are defined as the Euclidean distance between the audio features extracted via a VGG-like network \cite{133} pre-trained on AudioSet \cite{103}. In case the retrieved musical sample has the same genre as the given dance style, the segment is considered to be genre accurate. The genre accuracy is then calculated by $S_{c}/S_{t}$, where $S_{c}$ counts the number of genre accurate segments and $S_{t}$ is the total number of segments from the testing split. Note that the calculation of this objective metric requires the annotations of dance and music genres. Liang et al. \cite{77} assessed \textbf{Genre Accurate Distance (GAD)} and defined it as the average distance between genre accurate music features and the nearest GT music features.

\textbf{Video-Music CLIP Precision (VMCP)} \cite{12} is proposed to measure the video-music correspondence. A video-music CLIP model is first pretrained for this retrieval-based metric similar to \cite{141}. Given a generated music audio, VMCP is calculated as the top-K retrieval accuracy from a pool of M candidate videos using the CLIP model. Specifically, the cosine similarity scores between the generated sample and its condition video along with M-1 random sampled videos are ranked. The model is evaluated using all generated samples and computes the success retrieval rate as the final precision score. Zhuo et al. \cite{12} set M = 70, K = 5, 10, 20 in their experiments.

\textbf{Music Retrieval Precision.} Li et al. \cite{59} proposed a new metric named music retrieval to measure the music-video consistency. Given a piece of generated music and the ground truth music of its condition video, they randomly select M-1 pieces of music as distractors. Musicnn \cite{142} is used to extract music features for each generated item. If the ground-truth music ranks in the top-K place, then they consider it a successful retrieval. All generated samples are used to calculate the successful retrieval rate as the final precision score P@K. They set M = 64, K = 5, 10, 20. Since the ground truth music is related to the given video, the proposed retrieval precision metric is able to measure how well the generated music aligns with the given video. This metric aligns with the idea of VMCP, except that it calculates the similarity between music samples, while VMCP calculates the similarity between music and videos.

\textbf{Imagebind Score (IS)} \cite{4,15,143,144} assesses to what extent the generated music aligns with the videos. Specifically, it assesses the semantic relevance by calculating the cosine similarity of the embeddings for both audio and video. Despite the fact that Imagebind extends the CLIP to six modalities, only the branches of audio and vision are used. It is important to note that ImageBind is not specifically trained on music data. \textcolor{black}{Similarly, the LanguageBind \cite{183} can also project video and music into a unified textual latent space and the cosine distance between embeddings is calculated to produce the \textbf{LanguageBind (LB) score}.} Li et al. \cite{82} computed the average cosine similarity between video representations and music representations over the number of video frames. Furthermore, Liu et al. \cite{65} calculated the similarity scores for ranking purposes to obtain \textbf{ImageBind Ranking (IB Rank)}. \textcolor{black}{\textbf{CLAPScore} \cite{160} can also be used to measure the alignment between video-mapped musical descriptions and generated music.}

\textcolor{black}{Several metrics mentioned above share a common idea of computing cosine similarity to measure consistency either within a modality or across modalities, such as MCC, IS, LB Score, and CLAPScore. Specifically, MCC measures the similarity between generated and ground-truth music, focusing more on content fidelity within the music itself, whereas IS and others compute the similarity between the video (or video-driven text) and the generated music, better reflecting the cross-model semantic alignment. Consequently, for V2M generation, the latter is generally a more suitable cross-modal alignment metric.}

\paragraph{Temporal Synchronization.}
\textbf{Beats Coverage Score (BCS)} \cite{75,54,55} calculates the ratio of the overall generated beats to the total musical beats of the ground truth.

\textbf{Beats Hit Score (BHS)} \cite{75,54,55} measures the ratio of the aligned generated beats to the total musical beats of the ground truth.

Further, Yu et al. \cite{72} calculated the F1 scores of BCS and BHS as an integrated assessment and report the standard deviations of BCS and BHS (termed CSD and HSD, respectively) to evaluate generative stability. They also adapt BCS as the ratio of the aligned beats to the total beats from the generated music, for evaluating long-term soundtracks.

\textbf{Beat Alignment Score (BeatAlign)} \cite{75,77} is derived from music-to-dance research \cite{145,146}, is employed to assess motion-music correlation. This score quantifies the relationship between motion and music by computing the average distance between each kinematic beat (determined from the local minima of the kinetic velocity) and the nearest music beat (extracted using the Librosa library \cite{147}).

Additionally, Li et al. \cite{15} proposed a new metric that measures rhythm alignment by utilizing SceneDetect \cite{89} to identify frames with abrupt visual changes and detecting musical beats similar to the BHS. The metric is calculated as the recall rate of scene change frames at beats, allowing a tolerance of 0.1s, which is the ratio of the number of scene change frames that align with musical beats to the total number of scene change frames.

\textcolor{black}{\textbf{Audio-video alignment Score(AV-Align)} \cite{163,180} detects energy peaks in audio and video separately and measures how well they coincide. Audio peaks are obtained via an onset-detection algorithm \cite{181}, while video peaks are identified by computing the mean optical-flow magnitude per frame and locating rapid temporal changes. For each peak in one modality, the method validates whether a pick was also detected in the other modality within a three-frame temporal window and vice versa. The final alignment score, normalized by the total number of peaks, ranges from zero to one.}

\textbf{Tempo Difference (TD)} \cite{17} is the average L1 norm of tempo difference between generated and real music.

\textcolor{black}{
\textbf{Beats Intersection over Union (IoU)} \cite{172} metric measures the overlap, within a specified threshold, between the number of detected beats in generated music $B_{\text{syn}}$ and that in the ground truth $B_{\text{gt}}$, defined as:
$
B_{IoU} = 
\frac{B_{\text{gt}} \cap B_{\text{syn}}}{B_{\text{gt}} \cup B_{\text{syn}}}.
$
}

\textcolor{black}{\textbf{Cross-Modal Relevance (CMR) and Temporal Alignment (TA)} \cite{64,159,171} are proposed to evaluate the global and local (temporal) correspondence between video and music. Both metrics are derived from the same calculated value: the average of the diagonal elements of the cross-attention matrix $a(v, m)$ computed by a separately trained Evaluation Model. CMR serves as a global measure, with the Evaluation Model trained using InfoNCE Loss (contrastive learning) to maximize this relevance. TA functions as a local measure, achieved by training the model with MSE Loss to explicitly maximize the diagonal correlation.}

\subsection{Subjective}
Due to the perceptual nature of audio, human subjective assessment is widely regarded as the gold standard for evaluating generative models. As stated in \cite{27}, subjective evaluation includes methods such as the Turing test \cite{10}, binary/multiple-choice comparisons \cite{12,40,46,48,65}, and scoring \cite{38,54} based on subjective metrics defined by the researchers. Among these, the most fundamental and widely used metric is the Mean Opinion Score (MOS) \cite{149}. 

Gan et al. \cite{10} conducted a listening study on Amazon Mechanical Turk (AMT) to qualitatively compare the perceived quality of the generated music. Some studies \cite{17,51,54,55,70} conduct the Mean Opinion Scores (MOS) human test to assess the dance-music coherence and the general quality of the music samples. Kang et al. \cite{58} conducted a comprehensive listening test to rate music videos in view of the musical quality as well as the music-video alignment. Li et al. \cite{82} and Tong et al. \cite{172} conducted human evaluations with 5- and 10-point Likert scales, respectively, and report mean-opinion-scores (MOS-Q) audio quality and content alignment (MOS-A) with 95\% confidence intervals. 

The generated music is usually evaluated in terms of the following aspects \cite{4,12,58,59,77,90,160,163,165,173}: 
\begin{itemize}
\item Audio Quality: the perceptual clarity and fidelity of the audio;
\item Richness: the richness of the musical melody/harmony/rhythm/timbre;
\item Structure: the structure consistency of melody/rhythm;
\item Musicality: the aesthetic appeal of music independent of audio quality;
\item Overall Music Quality: the overall quality of the generated music independent of the video content;
\item Rhythmic/content/emotional/thematic/stylistic Correspondence: the correspondence of rhythm or content or emotion or theme or style between music and video;
\item Overall: a comprehensive evaluation that considers all above aspects of the music.
\end{itemize}

\subsection{Summary}
\textcolor{black}{These metrics complement each other by focusing on different aspects, jointly evaluating the quality of V2M generation methods. For instance, semantic alignment metrics measure cross-modal correspondence based on global information but may overlook fine-grained temporal correspondence between video and music, which can be addressed by temporal synchronization metrics, and vice versa.} Recently, Shi et al. \cite{148} proposed a versatile evaluation toolkit for speech, audio, and music called VERSA, which offers 63 metrics with 711 metric variations based on different configurations. In terms of music generation, VERSA can automatically calculate the aforementioned FAD, KLD, CLAP, Density, and Coverage, promoting fairness, reproducibility, and meaningful insights across studies. However, VERSA focuses more on text-based external resources, such as transcriptions and captions, lacking dedicated cross-modal metrics that assess semantic and rhythmic alignment between video and music.

\section{CHALLENGES}
The field of video-to-music generation holds immense potential for interdisciplinary innovation, bridging the realms of computer vision, audio processing, and artificial intelligence. As technology advances, this domain is poised to revolutionize industries such as entertainment, education, and virtual reality, offering immersive experiences where music dynamically adapts to visual narratives. Moreover, the fusion of aesthetic principles and cross-cultural musical traditions could lead to the development of universally resonant soundtracks, enriching global storytelling. According to the above survey, there has been a growing body of research on V2M generation in the past few years, with significant progress made in generating music that semantically and temporally aligns with a given video. However, several challenges and unresolved issues still persist as follows.

\textbf{Quality.} This challenge pertains to both the musicality of the generated music and its alignment quality with the video. Achieving an optimal trade-off between creative freedom and synchronization fidelity remains a critical hurdle. Striking the right balance is essential to ensure the music is engaging and harmonizes seamlessly with the video's rhythm and semantics. Note that the quality here does not include sound quality.

\textcolor{black}{
\textbf{Narrative-level Alignment}. A key long-term challenge is achieving narrative-level alignment, which requires models to move beyond simple beat or mood synchronization and instead follow the evolving narrative arc of a video. Current V2M systems lack mechanisms to understand high-level story progression, such as tension buildup, emotional turning points, or character-driven shifts, resulting in music that fails to capture global narrative flow. Advancing toward narrative-aware generation will require richer video representations, long-sequence modeling, and new learning objectives that encode story structure rather than short-term temporal patterns.
}

\textbf{External control. 1) Genre Control.} The genre of music generated by specialized models largely depends on the dataset used for training. While pretrained text-to-music models provide greater flexibility in controlling the global attributes of generated music, they often face challenges in maintaining temporal synchronization. This is primarily because accurately capturing time-varying rhythm characteristics through brief textual descriptions remains difficult. \textbf{2) Emotion Control.} Current V2M generation methods have largely overlooked the critical aspect of capturing and conveying emotional content, likely due to the scarcity of video-music datasets with detailed emotion annotations. While a few approaches attempt to extract emotions from videos, they suffer from inaccuracies in capturing and representing the intended emotions. Like genre control, pretrained text-to-music models can be employed to guide the emotional tone of the generated music. However, these methods face similar challenge that maps varying emotions into text embeddings.

\textcolor{black}{
\textbf{Fine-grained Editing}. In certain cases, users may be dissatisfied with only a specific portion of the generated music, such as a particular instrument or a few seconds of the track. In these situations, there is no need to regenerate the entire soundtrack but only modify the unsatisfactory parts. However, current V2M generation methods do not yet support such fine-grained editing and control. While music editing techniques do exist, in the context of V2M generation, it is essential that any edits maintain alignment and correspondence with the video content.
}

\textbf{Unified Framework.} Current approaches employ distinct feature extraction methods for human-centric and general videos, each emphasizing different aspects of video characteristics. For example, rhythm is commonly derived from human body movements in human-centric videos. While for general videos, rhythm is typically inferred from scene transitions or visual pacing. A unified framework that can effectively address both scenarios in a consistent and cohesive manner has yet to be developed. \textcolor{black}{Although Liu et al. \cite{170} claimed that their frame-wise dynamic feature encoding can address temporal synchronization across different scenarios, their experiments lack explicit evaluation of temporal alignment, leaving this claim to be further verified.}

\textbf{Personalization.} Personalized V2M generation remains a largely uncharted area. Current methods do not adequately account for individual preferences, such as tailoring the music to a specific cultural background or to mimic the reference music provided by users. Personalization would require models to adapt to user-specific criteria without sacrificing quality or alignment.

\textbf{Joint Generation.} The development of a joint generation framework for video and music remains underexplored. Existing joint generation efforts have primarily focused on human-centric videos for V2M generation, neglecting other general videos. A closed-loop system integrating both video-to-music and music-to-video processes could enable mutual correction and optimization of cross-modal features. This fusion perspective highlights the need for future research to jointly optimize both directions and address challenges such as feature alignment consistency, temporal synchronization, and semantic coherence in bidirectional generation, fostering a deeper integration of cross-modal understanding and generation.

\textcolor{black}{
\textbf{Human-AI Co-creation}. Generative models should serve as collaborative tools for artists rather than fully autonomous systems, and the same holds for V2M generation. Realizing this vision relies on addressing the aforementioned challenges of controllability and editability. Future systems may support iterative and interactive workflows, such as offering draft variations and fine-grained controls, and enable bidirectional feedback, in which users refine AI outputs while the system adapts to creative intent, ultimately expanding the expressive space of V2M generation.
}

\textcolor{black}{\textbf{Standard and Unified Evaluation.} A standardized evaluation framework remains an open challenge in the generative domain. Unlike classification tasks, which can be assessed using well-defined metrics such as accuracy or F1 score, generative content is frequently evaluated using different metrics across distinct studies. Furthermore, the continued collection of in-the-wild data has precluded the establishment of a unified dataset, rendering comparisons between models trained on disparate datasets unreliable. The development of standardized benchmarks and evaluation protocols such as MARBLE \cite{154}, which provides a benchmark for various Music Information Retrieval (MIR) tasks, would substantially alleviate this issue.}

\textbf{Non-instrumental-only Music.} To the best of our knowledge, existing V2M generation methods predominantly focus on instrumental music, lacking support for vocal components (e.g., lyrics and singing) and other auditory elements such as sound effects and speeches. Expanding these methods to incorporate non-instrumental elements poses unique challenges, particularly in achieving synchronization between vocal elements and the video's narrative structure and rhythmic patterns, as well as ensuring precise temporal alignment between visual events and sound effects.

\textcolor{black}{\textbf{Volumetric Video.} To the best of our knowledge, existing video-to-music generation studies have rarely explored the generation of spatial or stereo background music for volumetric videos. Stereo or multi-channel audio can provide a sense of spatial presence and directionality, which is crucial for enhancing immersion when viewing 3D or immersive content. Future research could investigate how to leverage the 3D structure, viewpoint changes, and dynamic scene information of volumetric videos to generate background music that aligns with the spatial characteristics, thereby enhancing realism and the overall immersive experience.}

\section{BROADER IMPACT}
The development of video-to-music generative models hold significant potential as foundational tools for content creators. While the primary goal is to empower creators in their creative pursuits rather than to replace human creativity, it is crucial to develop and implement these models with thoughtful consideration of the values, well-being, and impact on creators, their communities, and society at large.

\subsection{Ethical Concerns}
A critical concern lies in the inherent biases and patterns that generative models learn from their training data. In the video-to-music generation discussed in this paper, the model may propagate latent biases embedded within the video and music corpora. These biases often manifest in subtle and unpredictable ways, making them difficult to detect, and existing evaluation benchmarks may not fully capture these issues. For instance, the significant imbalances in music style distribution results in the generated music being limited to only a few genres. These issues extend to learned audiovisual associations, which may reinforce stereotypes, such as linking specific video content (e.g., people, body movements/dance styles, locations, objects) to a narrow range of musical styles, or trivializing the relationship between video narratives and audio outputs. Consequently, machine learning fairness testing is essential to understand the likelihood of such patterns emerging in specific models and to implement effective interventions \cite{78}.

In addition, there is a possibility that generated music may unintentionally resemble existing compositions, raising concerns about originality and ownership. Ensuring that AI-generated music adheres to copyright laws and respects the intellectual property of musicians remains a critical challenge. Implementing robust mechanisms, such as dataset curation, and automated similarity detection, could help mitigate these risks. Additionally, the legal framework surrounding AI-generated music needs further refinement to address the evolving landscape of digital content creation.

\subsection{Real-World Applications}
Technical advancements in video-to-music generation research have propelled its real-world applications, which are vast and transformative. \textcolor{black}{For example, Elevenlabs\footnotemark[8] allows users to upload any video (max 100MB) and its AI Music Generator composes an audio track that matches the tone, pace, and emotion of the visuals. No background in editing or music needed. The users can also contact their sales team to discuss a custom plan for high-volume use cases or to license Eleven Music for film, television, and video games. Suno Team newly proposed an app named Suno Scenes\footnotemark[7] that takes photos and videos and creates a unique song, enabling creators worldwide to inspire new musical content from captured memories and visual media. However, this app is available only on iOS. Beatoven.ai\footnote{https://www.beatoven.ai/} is an AI music generator that supports multimodal prompting to generate background music tailored to video content. Adobe Firefly\footnotemark[9] composes music for videos by analyzing videos and offering structured text prompts for text-to-music generation, typically following a template such as "a \textit{vibes} song, with \textit{genres} style, for a \textit{purpose}," allowing flexible soundtrack customization.  TianPuYue\footnote{https://www.tianpuyue.cn/}, launched by Quwan Network Technology Company, allows users to upload 10-60s videos and generate soundtracks.}  Below are some key domains where this technology could have a profound impact: 
\begin{itemize}
\item Film and Television: AI-generated music could dynamically adapt to visual content, reducing production costs while maintaining high-quality compositions. 
\item Game Development: Personalized and responsive soundtracks could be created based on in-game events, enhancing player immersion. 
\item Virtual Reality (VR) and Augmented Reality (AR): Real-time music generation could adapt to user interactions and environmental cues, making virtual experiences more engaging. 
\item Social Media and Content Creation: Video creators on platforms like YouTube and TikTok could use AI-generated music to match their visuals seamlessly without relying on expensive licensing agreements.
\end{itemize}

In tandem with algorithmic advancements, we call for future research to focus on 1) understanding and mitigating the risks associated with biases inherited from training data; 2) Enhancing models' ability to understand and respect cultural and social contexts, potentially through collaboration with experts in musicology and cultural studies; 3) Establishing clearer guidelines and frameworks for copyright and originality in AI-generated music, ensuring that musicians' rights are protected; 4) Involving content creators and end-users in the design and evaluation of these algorithms to ensure they meet real-world needs and ethical standards. We emphasize that these issues are as important and valuable as algorithmic progress itself, yet they are often overshadowed by the allure of technological advancements. Addressing these challenges is crucial to ensuring that generative models contribute positively to creative practices and societal well-being.

\section{CONCLUSION}
This paper presents a comprehensive survey of video-to-music generation using generative AI techniques. The focus of this paper is on generation algorithms, which typically consist of three key components: conditioning input construction, conditioning mechanism, and music generation. The design of these components varies according to the distinctive characteristics of the video and music modalities. Therefore, we begin by providing a fine-grained classification and analysis of both video and music modalities. Subsequently, we delve into the design schemes and strategic choices employed in existing studies for each of these three components. This structured organization allows researchers to reference methods applicable to similar video scenarios or music output formats, while facilitating targeted improvements to individual components and the flexible combination of different modules to develop new approaches. Additionally, we introduce available video-music datasets and evaluation metrics, while finally highlighting some challenges in existing methods. We hope this survey can help researchers quickly grasp the current state of the V2M generation field and guide further research from both data and algorithmic perspectives.

Video-to-music generation has advanced multimodal generation technologies in the integration of music and visual arts, fostering interdisciplinary collaboration and innovation. As V2M generation technologies continue to advance, they are expected to be widely applied in industries such as film production, video games, and virtual reality, potentially reshaping traditional approaches to music and art creation. While this survey focuses on V2M generation, it also provides valuable insights for other cross-modal or multimodal generation tasks, with significant academic, applicative, and societal implications.


\bibliographystyle{ACM-Reference-Format}
\bibliography{ref}

\end{document}